\documentclass[a4paper,12pt]{article}
\pdfoutput=1

\usepackage[a4paper,left=80pt,right=80pt,top=85pt,bottom=93pt]{geometry}
\usepackage[latin1]{inputenc}
\usepackage{dsfont}
\usepackage{amsfonts,amsmath,amssymb}
\usepackage{amsthm}
\usepackage{graphicx}
\usepackage[small,bf]{caption}
\usepackage{subcaption}
\usepackage{booktabs}
\usepackage[numbers, sort&compress]{natbib}
\allowdisplaybreaks[1]
\usepackage{color}
\usepackage{ulem}
\usepackage{tikz}
\usepackage{multirow}
\usetikzlibrary{arrows}
\usepackage{xspace}
\usepackage{enumitem}

\usepackage[pagebackref=false,colorlinks=false]{hyperref}

\def\bal#1\eal{\begin{align}#1\end{align}}
\newcommand{\be}{\begin{equation}}
\newcommand{\ee}{\end{equation}}
\newcommand{\bea}{\begin{eqnarray}}
\newcommand{\eea}{\end{eqnarray}}
\newcommand{\besub}{\begin{subequations}}
\newcommand{\eesub}{\end{subequations}}
\newcommand{\ba}{\begin{array}}
\newcommand{\ea}{\end{array}}
\newcommand{\bi}{\begin{itemize}}
\newcommand{\ei}{\end{itemize}}
\newcommand{\nn}{\nonumber}

\newcommand{\GeV}{{\rm GeV}\xspace}

\newcommand{\Mcal}{{\cal M}}
\newcommand{\Ocal}{\ensuremath{\mathcal O}}
\newcommand{\Lcal}{{\cal L}}

\newcommand{\hc}{{\textrm{h.c.}}}

\begin{document}

\begin{titlepage}

\flushright{ 
LPT-Orsay-16-63
\\
HIP-2016-31/TH}
 
\vspace*{0.7cm}

\begin{center}
{\LARGE 
{\bf
Multicomponent Dark Matter\\ from Gauge Symmetry
}}
\\
[1.0cm]

{
{\bf
Giorgio Arcadi$^{1,2}$, Christian Gross$^{3}$, Oleg Lebedev$^{3}$,\\ Yann Mambrini$^{1}$, Stefan Pokorski$^{4}$, Takashi Toma$^{1}$
}}
\end{center}
%\addtocounter{footnote}{-3}
\vspace*{0.3cm}

\centering{
$^{1}$ 
\it{
Laboratoire de Physique Th\'eorique, CNRS -- UMR 8627, \\
Universit\'e de Paris-Saclay 11, F-91405 Orsay Cedex, France
}

$^{2}$ 
\it{
Max Planck Institut f\"{u}r Kernphysik, Saupfercheckweg 1,\\
D-69117 Heidelberg, Germany
}

$^{3}$ 
\it{Department of Physics and Helsinki Institute of Physics, \\
Gustaf H\"allstr\"omin katu 2, FI-00014 Helsinki, Finland
}

$^{4}$ 
\it{Institute of Theoretical Physics, University of Warsaw, \\
Pasteura 5, PL-02-093 Warsaw, Poland
}
}

\vspace*{1.0cm}

\begin{abstract}
The composition of Dark Matter (DM) remains an important open question. The current data do not distinguish between single-- and multi--component DM, while in theory constructions it is often assumed that DM is composed of a single field. In this work, we study a hidden sector which naturally entails multicomponent DM consisting of spin-1 and spin-0 states. This UV complete set-up is based on SU(3) hidden gauge symmetry with the minimal scalar field content to break it spontaneously. The presence of multiple DM components is a result of a residual $Z_2 \times Z'_2$ symmetry which is part of an unbroken global $U(1) \times Z_2^\prime$ inherent in the Yang--Mills systems. We find that the model exhibits various parametric regimes with drastically different DM detection prospects. In particular, we find that the direct detection cross section is much suppressed in large regions of parameter space as long as the Standard Model Higgs mixes predominantly with a single scalar from the hidden sector. The resulting scattering rate is often beyond the level of sensitivity of XENON1T, while still being consistent with the thermal WIMP paradigm.

\end{abstract}

%\today

\end{titlepage}
\newpage

\tableofcontents

%=========================================================================
%=========================================================================
\section{Introduction}
%=========================================================================
%=========================================================================

The existence of a Dark Matter (DM) component of the Universe is confirmed by several astrophysical and cosmological probes, e.g.~the CMB~\cite{PLANCK} and structure formation. 
Particle physics solutions to the DM problem mostly rely on the existence of a new particle that is stable on cosmological scales thanks to a symmetry, and with weak enough interactions with Standard Model (SM) states to evade constraints from direct and indirect searches. 
From the model-building point of view, these requirements are naturally satisfied by 
 a ``hidden'' sector whose states are singlets with respect to the SM symmetry group. 
In this kind of a setup, stable particles of the hidden sector are Dark Matter candidates.

Despite the different symmetry groups acting on the visible and hidden sectors, renormalizable interactions can arise among them. 
The dimension--4 operators relevant to our study are obtained by combining the gauge invariant dimension two terms $H^{\dagger}H$ and $B^{\mu \nu}$ with similar dimension-two operators formed by the states of the hidden sector.
The strength of such ``portal'' interactions can be sufficient to bring the visible and
hidden sectors in thermal equilibrium in the Early Universe and realize the WIMP paradigm.
In addition, the DM candidates retain interactions with the visible particles at present times, which could be within 
 the reach of the current and future searches for new particles. 

In the simplest models of this type, the hidden sector is populated (effectively) by a single field
which constitutes DM.
The lowest order Higgs portal operators mediating interactions between the DM and the SM states then read $H^{\dagger} H |\chi|^2$ or $H^{\dagger} H V^\mu V_\mu$, with $\chi$ and $V^\mu$ being a scalar and a vector DM candidate, respectively, see e.g.~\cite{Silveira:1985rk,Patt:2006fw,MarchRussell:2008yu,Andreas:2008xy,Hambye:2008bq, 
Kanemura:2010sh,Lebedev:2011iq,Djouadi:2011aa}.\footnote{An analogous fermionic Higgs portal interaction~\cite{Kim:2006af,LopezHonorez:2012kv} is dimension-5. Note also that
even though naive dimension counting gives 4 for $H^{\dagger} H V^\mu V_\mu$, it actually originates from a dim-6 operator~\cite{Lebedev:2011iq}. See also related analyses in \cite{Farzan:2012hh,Baek:2012se,Duch:2015jta}.}
These simple set--ups are currently under pressure from constantly improving
 experimental constraints.
 Indeed, by crossing symmetry arguments, there is a relation between the DM pair annihilation cross-section at freeze-out, responsible for the relic density, and the processes potentially responsible for detection signals, such as scattering on nuclei, probed by Direct Detection experiments, or production at colliders.
 Null results of the latter then rule out large portions of the parameter space favoured
 by the thermal WIMP paradigm~\cite{LUX,Tan:2016zwf}.
 If the s-wave DM annihilation cross section remains substantial at present times,
 the model can also be probed by Indirect Detection experiments.
These considerations motivate exploration of richer hidden sector structures.
For example, additional annihilation channels into dark sector states can deplete the DM relic density without changing its interactions with the visible sector 
 while satisfying the experimental constraints~\cite{Belanger:2011ww,Dudas:2014ixa}.

More generally, there is no a priori reason for the DM of the Universe to be composed of a single field.
Multi-component DM frameworks, with two or more particles contributing a non-negligible fraction to the total relic density $\Omega_{\rm DM, \rm tot}h^2\approx 0.12$,
 offer interesting perspectives. The relation between the annihilation cross section and the current detection signals has to be properly reconsidered.
Some work in this direction has been carried out in~\cite{Profumo:2009tb,Dienes:2012cf},
where 
 the discovery potential of the current and future experimental facilities and the capability of discriminating multicomponent DM from single--component DM have been studied. We note that multicomponent
 DM emerges in various particle physics models (see e.g. \cite{Esch:2014jpa,Boddy:2014yra, Klasen:2016qux,DiFranzo:2016uzc}).

In this work, we will investigate multicomponent DM emerging from a hidden sector endowed with gauge symmetry.
Such systems enjoy natural discrete symmetries which can act as DM stabilizers.
Indeed, it was noted in \cite{Hambye:2008bq} and 
detailed in \cite{Lebedev:2011iq} that a hidden sector consisting of a U(1) gauge field $A_{\mu}$ and a single complex scalar 
which breaks the symmetry spontaneously has the symmetry
\begin{equation}
Z_2~:~~~A_{\mu} \rightarrow - A_{\mu} \;.
\end{equation}
As a result, the massive vector field $A_\mu$ is stable and can constitute DM.
This idea generalizes to non--Abelian gauge symmetries as well.
In particular, hidden SU(N) sectors with a minimal matter content 
necessary to break the gauge symmetry completely, that is $N-1$ scalar $N$-plets,
are endowed with a $Z_2 \times Z_2^\prime$ symmetry~\cite{Gross:2015cwa}\footnote{This assumes {\it CP}-symmetry of the hidden sector scalar potential.}
 \begin{eqnarray}
&& Z_2~:~~~A_{\mu}^a \rightarrow (-1)^{n_a} A_{\mu}^a \;, \nonumber\\
&& Z_2^\prime ~:~~~A_{\mu}^a \rightarrow (-1)^{n_a'} A_{\mu}^a \;,
\label{z2z2}
\end{eqnarray}
where $A_{\mu}^a$ is the gauge field, $a$ is the adjoint group index
and $n_a,n_a'$ take on values 0,1 depending on $a$. One of the $Z_2$'s
corresponds to complex conjugation of the SU(N) group elements, while the other is a gauge transformation. 
In the SU(2) case,
the symmetry enlarges in fact to custodial SO(3) \cite{Hambye:2008bq}
(see also~\cite{Karam:2015jta,Khoze:2016zfi}), while for larger groups 
it is part of a global unbroken $ U(1) \times Z_2^\prime$.
Since the SM fields are neutral under the above symmetries, $A_{\mu}^a$
cannot decay into the visible sector particles and therefore can
constitute dark matter. Note that since the symmetry is 
$Z_2 \times Z_2^\prime$, one expects at least $two$ different DM components, in contrast to traditional $Z_2$--invariant WIMP models. 
The exact composition depends on the details of the spectrum. In particular, the SU(3) example studied in \cite{Gross:2015cwa} has
only vector DM with two components being degenerate in mass and the 
third one being somewhat lighter.
These states interact with the visible sector
through the Higgs portal operators.
 A related study has recently appeared in \cite{Karam:2016rsz}.
 
 In our current study, we explore a qualitatively different case of {\it mixed spin} DM,
 that is containing both spin 1 and spin 0 components.
 We employ the 
 model of \cite{Gross:2015cwa} in a different parametric regime, 
 where a stable pseudoscalar is lighter than the gauge field with the same $Z_2 \times Z_2^\prime$ quantum numbers. In this case, the pseudoscalar as well as the gauge fields with distinct $Z_2 \times Z_2^\prime$ quantum numbers constitute DM. The resulting phenomenology
 is very different from that of \cite{Gross:2015cwa}. In particular,
we find that there are substantial regions of parameter space where the direct detection cross section is suppressed. 
 
 We stress that although we study a specific model of multicomponent DM, many of the results presented here are of general relevance.
 In particular, depending on the composition of DM, the direct detection signal strength varies drastically, over orders of magnitude, and is often consistent with thermal relic DM abundance. 
Such behaviour is specific to more complicated hidden sectors within
our framework and reflects the possibility that common models may
oversimplify the DM properties.

One of the novel aspects of our study is that multicomponent DM 
is a natural consequence of our UV--complete framework, due to $Z_2 \times Z_2^\prime$ being part of the Yang--Mills symmetries. This
is in contrast to more conventional models where the two DM components have different origins such as the mixed axion--neutralino 
DM scenario~\cite{Bae:2015rra,Badziak:2015qca}.
Consequently, the contributions of the components to the total DM
density are controlled by a set of the UV parameters. In our study, much emphasis will be given to the analysis of the DM production processes 
(as opposed to the approach of~\cite{Profumo:2009tb,Dienes:2012cf}).
We solve numerically the coupled Boltzmann equations and calculate the individual relic abundances as a function of the parameters of the model. The composition of DM can be very different in different parameter regions and in some of them both DM components give comparable contributions. 
We then study the Direct Detection constraints and observe
an interesting effect. As long as the SM Higgs mixes predominantly
with one of the hidden scalar fields, the direct detection is highly
suppressed in the parameter regions where DM is mostly spin-0. 
It can be so small that even future detectors like XENON1T~\cite{Aprile:2015uzo} will not be able to probe it. This is one 
of the main results of our study.
 
 The paper is structured as follows.
The model is introduced in Section~2. 
Section~3 is devoted to a detailed discussion of the relic DM density calculations  and the Direct Detection limits. We also comment on
the possibility of detecting one of the components through Indirect Detection.
Our results are summarized in Section~4.

%=========================================================================
%=========================================================================
\section{The SU(3) hidden sector model}
%=========================================================================
%=========================================================================

The purpose of this section is to briefly summarise our model, mostly following Ref.~\cite{Gross:2015cwa}.
The hidden sector of the model is endowed with SU(3) gauge symmetry,
which is broken spontaneously (to nothing) by two hidden triplets
$\phi_1$ and $\phi_2$. This is the minimal setup that allows one to 
make all the SU(3) gauge fields massive.

The Lagrangian of the model is
\be
\Lcal_{\rm SM} + \Lcal_{\rm portal} + \Lcal_{\rm hidden} \;,
\ee
where 
\besub
\bal
-\Lcal_{\rm SM} &\supset V_{\rm SM} =\frac{\lambda_{H}}{2} |H|^4+m_{H}^2 |H|^2 \;,
 \\
-\Lcal_{\rm portal} &= V_{\rm portal}= \lambda_{H11} \, |H|^2 |\phi_1|^2 + \lambda_{H22} \, |H|^2 | \phi_2|^2 - ( \lambda_{H12} \, |H|^2 \phi_1^\dagger \phi_2 +\hc)\;,
 \\
\Lcal_{\rm hidden} &= - \frac12 \textrm{tr} \{G_{\mu \nu} G^{\mu \nu}\} + |D_\mu \phi_1|^2 + |D_\mu \phi_2|^2 -V_{\rm hidden} \,.
\eal
\eesub
Here, $G_{\mu \nu}=\partial_\mu A_\nu - \partial_\nu A_\mu + i \tilde g [A_\mu,A_\nu]$ is the field strength tensor of the SU(3) gauge fields $A_\mu^a$ with gauge coupling $\tilde g$, $D_\mu \phi_i = \partial_\mu \phi_i + i \tilde g A_{\mu} \phi_i$ is the covariant derivative of $\phi_i$, $H$ is the Higgs doublet, which in the unitary gauge can be written as $H^T= (0,v+h)/\sqrt{2}$, and the most general renormalisable hidden sector scalar potential is given by 
\bal
V_{\rm hidden}(\phi_1,\phi_2) &=
m_{11}^2 |\phi_1|^2
+ m_{22}^2 |\phi_2|^2
- ( m_{12}^2 \phi_1^\dagger \phi_2 + \hc )
\nn \\ 
& 
+ \frac{\lambda_1}{2} |\phi_1|^4
+ \frac{\lambda_2}{2} |\phi_2|^4
+ \lambda_3 |\phi_1|^2 |\phi_2|^2
+ \lambda_4 | \phi_1^\dagger\phi_2 |^2
\nn \\ 
& 
+ \left[
\frac{ \lambda_5}{2} ( \phi_1^\dagger\phi_2 )^2
+ \lambda_6 |\phi_1|^2
( \phi_1^\dagger\phi_2)
+ \lambda_7 |\phi_2|^2
( \phi_1^\dagger\phi_2 )
+ \hc \right] \,.
\label{V}
\eal
The fields $\phi_1$ and $\phi_2$ are responsible for the spontaneous breaking of the hidden SU(3) symmetry. In the unitary gauge, they can be written as
\be \label{unitarygauge}
\phi_1={1\over \sqrt{2}} \,
\left( \begin{array}{c}
0\\0\\v_1+\varphi_1
\end{array} \right) \,,
\quad 
\phi_2= {1 \over \sqrt{2}}\,
\left( \begin{array}{c}
0\\v_2+\varphi_2\\v_3+\varphi_3 + i \varphi_4
\end{array} \right) ~,
\ee
where the $v_i$ are real VEVs and $\varphi_{i}$ are real scalar fields. 
Here we assume the $CP$ symmetry in the scalar sector, i.e. that the couplings are real and $\varphi_4$ attains no VEV.
 As a consequence there is no mixing between the {\it CP}-even scalar fields $\varphi_{1-3}$ and the {\it CP}-odd scalar $\varphi_4$. 
This allows for the possibility that $\varphi_4$ is stable.

A minor technical complication that occurs in this model is that the quadratic part of the Lagrangian is not diagonal, due to the mixing terms
\begin{equation}
\label{eq:gauge_scalar_mixing}
\mathcal{L}\supset \frac{\tilde{g}v_2}{2}A^{\mu\, 6}\partial_\mu \varphi_4-\frac{\tilde{g}v_3}{\sqrt{3}}A^{\mu\, 8}\partial_\mu \varphi_4+\frac{\tilde{g}v_3}{2}A^{\mu\, 7}\partial_\mu \varphi_2-\frac{\tilde{g}v_2}{2}A^{\mu\, 7}\partial_\mu \varphi_3 \,.
\end{equation}
These terms of the form $\kappa_{ai} A_\mu^a \partial^\mu \varphi^i$ can be removed by the transformation 
\begin{equation}
\label{eq:gauge_transform}
A^a_\mu \rightarrow \tilde{A}^a_\mu = A^a_\mu+\partial_\mu Y^a, \quad \textrm{with} \quad Y^a \equiv (\mathcal{M})^{-1}_{ab} \kappa_{bi} \varphi_i \,,
\end{equation}
where $\mathcal{M}$ is the mass matrix of the hidden gauge bosons.
This leaves $\mathcal{M}$ unchanged.
After the above transformation the kinetic terms of the $\varphi_i$ are not canonically normalised anymore so that a further transformation
\be \label{normalizescalars}
\varphi_i \rightarrow \tilde \varphi^i = \omega_{ik} \varphi^k \;, \quad \textrm{where} \quad (\omega^T \omega)_{ij}\equiv \delta_{ij} - \kappa^T_{ia}\Mcal^{-1}_{ab}\kappa_{bj} \,,
\ee
is needed to make the quadratic part of the Lagrangian canonically normalised.
To stress its special role as a DM candidate, we relabel
\be
\chi \equiv \tilde{\varphi}_4 \,.
\ee

To simplify the analysis, 
in the rest of the paper we assume that the couplings $m_{12}^2,\lambda_{H12},\lambda_6,\lambda_7$ in the scalar potential, as well as the VEV $v_3$, are small but non-vanishing. 
If they did vanish, the system would attain an additional unwanted
$Z_2$ symmetry $\phi_2 \rightarrow -\phi_2$, which would lead to 
extra stable particles and change the phenomenology of the model. 
In the limit of small $v_3$, the only gauge-scalar mixing terms are $A^{\mu\,6}\partial_\mu \varphi_4$ and $A^{\mu\,7}\partial_\mu \varphi_3$ so that $\tilde \varphi_1 \simeq \varphi_1$ and $\tilde \varphi_2 \simeq \varphi_2$.

The mass matrix for the (pseudo)scalar fields reads:
\begin{equation}
-\mathcal{L}\supset \frac{1}{2} \Phi^T m_{\mathrm{{\it CP}-even}}^2 \Phi+\frac{1}{4}\left(\lambda_4-\lambda_5\right) (v_1^2+v_2^2) \chi^2,
\end{equation}
where $\Phi=(h,\varphi_1,\varphi_2,\tilde{\varphi_3})^T$. In the limit $v_3 \ll v_1,v_2$, we get
\begin{equation} \label{massmat}
m_{\mathrm{{\it CP}-even}}^2 = \left(
\begin{array}{cccc}
\lambda_Hv^2 & \lambda_{H11}vv_1 & \lambda_{H22}vv_2 & 0\\
\lambda_{H11}vv_1 & \lambda_1v_1^2 & \lambda_3v_1v_2 & 0\\
\lambda_{H22}vv_2 & \lambda_3v_1v_2 & \lambda_2v_2^2 & 0\\
0 & 0 & 0 & (\lambda_4+\lambda_5)(v_1^2+v_2^2)/2
\end{array}
\right).
\end{equation}
We see that $\tilde{\varphi}_3$ does not mix with the other states and is a mass eigenstate.
The other mass eigenstates are obtained by diagonalising the upper $3 \times 3$ sub-matrix. 
For further simplification, we will assume that the (1,2) and (2,3) entries of $m_{\mathrm{{\it CP}-even}}^2$ are much smaller than the other matrix elements, which can be achieved with sufficiently small $\lambda_{H11}$ and $\lambda_3$.
Then $ \varphi_1$ is approximately a mass eigenstate, which we call ${\cal H}$ (to be consistent with the notation in Ref.~\cite{Gross:2015cwa}), and $m_{{\cal H}}^2=\lambda_1 v_1^2$.
The other two mass eigenstates are\footnote{We will often abbreviate $s_\theta \equiv \sin \theta$ and $c_\theta \equiv \cos \theta$.}
\begin{align}
& h_1\simeq c_\theta h - s_\theta \varphi_2 ~,\nonumber\\
& h_2\simeq s_\theta h + c_\theta \varphi_2 ~,
\end{align}
with
\begin{align}
& m_{h_1,h_2}^2\simeq \frac{1}{2}\left(\lambda_2 v_2^2+\lambda_H v^2\right) \mp \frac{\lambda_2 v_2^2-\lambda_H^2 v^2}{2 c_{2\theta}}~, \nonumber\\
& \tan 2\theta \simeq \frac{2 \lambda_{H22} v v_2}{\lambda_2 v_2^2-\lambda_H^2 v^2} \,.
\end{align}

The eigenstate $h_1$ is identified with the 125 GeV Higgs boson and, consequently, its couplings are required to be SM-like. This translates into the requirement $s_\theta \lesssim 0.3$ (see e.g.~\cite{Falkowski:2015iwa}).

We now turn to the vectors.
In the limit $v_3 \ll v_1,v_2$ the vector sector is composed of 6 pure states which form 3 mass degenerate pairs with masses
\begin{equation}
\label{eq:vmasses_1}
m_{A^1}^2=m_{A^2}^2=\frac{\tilde{g}^2}{4}v_2^2,\quad
m_{A^4}^2=m_{A^5}^2=\frac{\tilde{g}^2}{4}v_1^2,\quad
m_{A^6}^2=m_{A^7}^2=\frac{\tilde{g}^2}{4}(v_1^2+v_2^2) \,,
\end{equation}
and two mixed eigenstates
\begin{align}
& A_\mu^{3\,'}=A_\mu^3 \cos\alpha + A_\mu^8 \sin\alpha ~,\nonumber\\
& A_\mu^{8\,'}=A_\mu^8 \cos\alpha - A_\mu^3 \sin\alpha~,
\end{align}
where\footnote{Note that this definition of $\alpha$ differs from that in Ref.~\cite{Gross:2015cwa} for $v_2^2 > 2 v_1^2$ by $\frac \pi 2$.
With the definition used here $A^{3\,'}$ is always the lightest vector.}
\be
\alpha=\left \{ 
\begin{array}{ll}
\frac12 \arctan \left( \frac{\sqrt{3}v_2^2}{2 v_1^2-v_2^2} \right)
&
\textrm{for} \quad v_2^2 \leq 2 v_1^2
\vspace{5pt}
\\
\frac12 \arctan \left( \frac{\sqrt{3}v_2^2}{2 v_1^2-v_2^2} \right) + \frac \pi 2
&
\textrm{for} \quad v_2^2 > 2 v_1^2 ~,
\end{array}
\right.
\ee
so that $\alpha \in (0^\circ, 60^\circ)$.
The masses are
\begin{align}
\label{eq:vmasses_2}
& m_{A^{3\,'}}^2=\frac{\tilde{g}^2 v_2^2}{4}\left(1-\frac{\tan \alpha}{\sqrt{3}}\right),\,\,\,\,\,m_{A^{8\,'}}^2=\frac{\tilde{g}^2 v_1^2}{3}\frac{1}{1-\frac{\tan \alpha}{\sqrt{3}}} \,.
\end{align}

Our setup enjoys a $Z_2 \times Z^{'}_2$ symmetry (cf. Eq.~(\ref{z2z2})). 
The $Z_2^\prime$ acts
as complex conjugation, which is an outer automorphism of SU(3),
while the $Z_2$ is a gauge transformation that acts non--trivially
only on the upper entry of the SU(3) triplets. They are inherent in
the Yang--Mills system and remain unbroken by interactions with matter in our minimal setting.

This discrete symmetry is in fact part of a global $U(1) \times Z_2^\prime$
preserved by the vacuum. 
The global U(1)
\begin{equation}
U= e^{ i \beta /3} \;{\rm diag} (e^{- i \beta} , 1 , 1)  
\end{equation}
is a subgroup of the SU(3) hidden gauge symmetry and acts on the gauge fields as $A_\mu \to U A_\mu U^\dagger$. 
This corresponds to $(A_{1,4},A_{2,5})\to (\cos \beta A_{1,4}- \sin \beta A_{2,5},\sin \beta A_{1,4} + \cos \beta A_{2,5})$ and leaves $A_{3,6,7,8}$ invariant.
The scalar sector Eq.~(\ref{V}) possesses an independent  global U(1)$^\prime$ symmetry 
$\phi_{1,2} \to  e^{i \gamma}\;\phi_{1,2}$. 
Since $U$ acts effectively as an overall phase transformation on the 
scalar fields of the form Eq.~(\ref{unitarygauge}), the vacuum  
preserves a combination of U(1) and U(1)$^\prime$. This symmetry ensures, for instance, that $m_{A_1}= m_{A_2}$ and $m_{A_4}= m_{A_5}$ 
(see \cite{Gross:2015cwa}).
   
Although the unbroken symmetry is $U(1) \times Z_2^\prime$, for our
purposes it suffices to consider its subgroup $Z_2 \times Z^{'}_2$.
The corresponding charges are given in Table~\ref{tab:dm}.
\begin{table}[h]
\begin{center}
\begin{tabular}{c|c|c}\hline
gauge eigenstates & mass eigenstates &
 $Z_2\times Z_2^{\prime}$
\\\hline\hline
$h,\varphi_{1-3},A_\mu^7$ &
$h_{1,2}, {\cal H}, {\tilde \varphi}_{3}, \tilde{A}_\mu^7$ & $(+,+)$\\
$A_\mu^1,A_\mu^4$ &
$A_\mu^1,A_\mu^4$ & $(-,-)$\\
$A_\mu^2,A_\mu^5$ &
$A_\mu^2,A_\mu^5$ & $(-,+)$\\
$\varphi_4,A_\mu^3,A_\mu^6,A_\mu^8$ &
$\chi,A_\mu^{\prime3},\tilde{A}_\mu^6,A_\mu^{\prime8}$ & $(+,-)$\\\hline
\end{tabular}
\caption{$Z_2\times Z_2^{\prime}$ charges of the scalars and hidden vectors.}
\label{tab:dm}
\end{center}
\end{table}

The lightest vector state is always $A^{3\,'}$. 
It is however not necessarily stable since $|D_\mu \phi_2|^2$ 
generates the coupling
\be
{\cal L} \supset (1+r) \frac{\tilde g}{\sqrt{3}} \sin \alpha \left( \chi A_\mu^{3\,'} \partial^\mu \tilde \varphi_3 - \chi \leftrightarrow \tilde \varphi_3 \right)~,
\ee
where
\begin{equation}
r \equiv v_2^2/v_1^2 \,,
\end{equation}
 allowing for the decay $A^{3\,'} \rightarrow \chi + {\tilde \varphi_3}   \rightarrow \chi + {\rm SM}$ if $m_{A^{3\,'}} > m_\chi$.
 Here ${\tilde \varphi_3}$ is produced off-shell and leads to the SM
 final states since 
 the coupling of ${\tilde \varphi_3}$ to $h_1, h_2$ is nonzero for $v_3 \neq 0$.

The masses of $A^{3\,'}$ and $\chi$ are related by
\be
\label{eq:A3_ratio}
\frac{m_{\chi}^2 }{m^2_{A'^3}}= \frac{\lambda_4-\lambda_5}{\tilde g^2} f(r) \,,
\quad
\textrm{with}
\quad
f(r)= \frac{3 (r+1)}{r+1-\sqrt{1+r(r-1)}} \,.
\ee
The decay $A^{3\,'}\rightarrow \chi$ + SM is thus kinematically open if $\lambda_4-\lambda_5 < \tilde g^2/f(r)$. 
For $r$ around unity, one has $f(r) = 6 + \Ocal((1-r)^2)$, while for $r \ll 1$, $f(r) \simeq 2/r+ \Ocal(1)$.
If one requires $\chi$ to be part of DM, relatively small $\tilde g$ necessitates therefore very small $\lambda_4-\lambda_5$.

In summary, our SU(3) hidden sector adds to the particle content 
the following states: 8 massive vector bosons, three scalars $h_2, {\cal H}, \tilde \varphi_3$ and one pseudo-scalar $\chi$. Given the charges under the $Z_2 \times Z_2^{'}$ symmetry and the mass relations of Eqs.~(\ref{eq:vmasses_1},\ref{eq:vmasses_2},\ref{eq:A3_ratio}), different states can contribute to Dark Matter. The options
are summarized in Table~\ref{tab:dm-patterns}.
In all cases, DM consists of 3 states. Since $A_1,A_2$ 
are degenerate in mass, one may introduce a formal analog of the $W^\pm$ 
bosons via the linear combinations $A_1 \pm i A_2$ even though 
$A_1,A_2$ have different parities. We find that 
such a redefinition facilitates numerical computations, in particular, what concerns the software Micromegas. This allows for the treatment
of $A_1,A_2$ as an effectively single (complex) DM component.
A similar redefinition can be applied to another mass degenerate pair
 $A_4,A_5$ .
\begin{table}[h]
\begin{center}
\begin{tabular}{c|c|c|c|c}\hline
& Case I & Case II & Case III & Case IV
\\\hline\hline
 parameter & $v_2<v_1$ & $v_2>v_1$ & $v_2<v_1$ & $v_2>v_1$\\
 choice & $\lambda_4-\lambda_5$ small & $\lambda_4-\lambda_5$ small & $\lambda_4-\lambda_5 \geq\mathcal{O}(1)$ & $\lambda_4-\lambda_5\geq\mathcal{O}(1)$
 \\\hline
 dark matter 
 & $A_{\mu}^1$,$A_{\mu}^2$,$\chi$
 & $A_{\mu}^4$,$A_{\mu}^5$,$\chi$
 & $A_{\mu}^1$,$A_{\mu}^2$,$A_{\mu}^{\prime3}$
 & $A_{\mu}^4$,$A_{\mu}^5$,$A_{\mu}^{\prime3}$
 \\\hline
\end{tabular}
\caption{ DM composition for different parameter choices
(cf. Eq.~(\ref{eq:A3_ratio})). }
\label{tab:dm-patterns}
\end{center}
\end{table}

The four possible cases can be understood as follows. For $v_2 < v_1$ ($v_1 < v_2$) the degenerate pair $A_{1,2}$ ($A_{4,5}$) is stable, because these are the lightest states with a given non--trivial $Z_2 \times Z_2^{'}$ charge. 
The only possible decay would be of the type $A_{1} \rightarrow A_2 A_3$ which is kinematically forbidden.
 The other stable state of the hidden sector is either $\chi$ or $A^{3\,'}$, depending on the value of $\lambda_4-\lambda_5$.
The purely vectorial DM case was studied in~\cite{Gross:2015cwa}.
 In this work, we will instead focus on case I, with mixed scalar--vector DM.

%=========================================================================
%=========================================================================
\section{Multicomponent Dark Matter Phenomenology}
%=========================================================================
%=========================================================================

Many of the important features of our model can be obtained by taking the limit
 $v_1 \gg v_2$. This reduces the number of states relevant to DM phenomenology to 
 the DM candidates $A_{1,2}$ and $\chi$, two mediators $h_1$ and $h_2$, and the state $A_3$ whose mass is between that of $A_{1,2}$ and $\chi$.
 We discuss this limit in the next subsection.
 Afterwards, we also consider the case $v_1\simeq v_2$ where all hidden states play a role and highlight the differences between these two limits.

%=========================================================================
\subsection{Case $v_1 \gg v_2$}
%=========================================================================

For $v_1 \gg v_2$, the mass scales of $A_{1,2}$,$A^{'}_{3}$ on one hand and $A_{4-7},A^{'}_8$ on the other hand are split, with the latter being higher by a factor of order $v_1 / v_2$.
The same happens in the scalar sector where the states ${\cal H}, {\tilde \varphi}_{3}$ are parametrically heavier than $h_{1,2}$. 
On the other hand, since we are interested in a relatively light $\chi$, we take a small enough value of $\lambda_4-\lambda_5$ to keep its mass below that of $A^{'}_{3}$ 
(cf. Eq.~(\ref{eq:A3_ratio})). In practice, $v_1/v_2 \simeq 3-5$ is sufficiently large
 to neglect the heavier states, while we take $v_1/v_2 =10$
 in our numerical studies.

For $v_1 \gg v_2 (\gg v_3)$,
the relevant for our purposes Lagrangian is given by
\begin{equation}
\label{eq:simple_lagrangian}
\mathcal{L}=\mathcal{L}_{\rm DM}+\mathcal{L}_{\scriptscriptstyle h\text{-}\rm SM\text{-}SM} +\mathcal{L}_{h\text{-}h\text{-}h} \;,
\end{equation}
where we neglect the $h^4$--type couplings which do not contribute significantly to the DM relic density computations. 
Here, the DM Lagrangian, containing the mass terms and the $h_1,h_2$
interaction terms, is
\bal
\mathcal{L}_{\rm DM} & = \frac12 m_{A^{}}^2 \sum_{a=1,2} A^a_\mu A^{a\mu}
- \frac12 m_{\chi}^2 \chi^2 
\label{DMDMh}
\\
&+ \left[ \frac{ \tilde g \, m_{A^{}}}2 \left(- h_1 s_\theta + h_2 c_\theta\right)
+ \frac{\tilde g^2}{8} \left(h_1^2 s^2_\theta - 2 h_1 h_2 s_\theta c_\theta + h_2^2 c^2_\theta \right) \right]
 \sum_{a=1,2} A^a_\mu A^{a\mu} \nn
\\
&+ \left[ \frac{\tilde{g} (1+r)}{2 m_A} \left(-h_1 s_\theta m_{h_1}^2 +h_2 c_\theta m_{h_2}^2 \right) 
- \frac 14 \left( \lambda_{\chi \chi 1 1} h_1^2 + 2 \lambda_{\chi \chi 1 2} h_1 h_2 + \lambda_{\chi \chi 22} h_2^2 \right) \right]
\chi^2 
\nn
\,,
\eal
where
\besub
\bal
\lambda_{\chi \chi 1 1} &=
(1+r) \frac{\tilde g} { 2 m_A v} s_\theta \left( c^3_\theta (m_{h_2}^2 - m_{h_1}^2) + \frac{\tilde g v} { 2 m_A} s_\theta (s^2_\theta m_{h_1}^2 + c^2_\theta m_{h_2}^2 ) \right)~,
\\
\lambda_{\chi \chi 1 2} &=
(1+r) \frac{\tilde g} { 2 m_A v} s_\theta c_\theta \left( s_\theta c_\theta (m_{h_2}^2 - m_{h_1}^2) - \frac{\tilde g v} { 2 m_A} (s^2_\theta m_{h_1}^2 + c^2_\theta m_{h_2}^2 ) \right)~,
\\
\lambda_{\chi \chi 2 2} &=
(1+r) \frac{\tilde g} { 2 m_A v} c_\theta \left( s^3_\theta (m_{h_2}^2 - m_{h_1}^2) + \frac{\tilde g v} { 2 m_A} c_\theta (s^2_\theta m_{h_1}^2 + c^2_\theta m_{h_2}^2 ) \right) \,.
\eal
\eesub
The couplings of $h_1$ and $h_2$ to SM matter are given by
\be
\mathcal{L}_{\scriptscriptstyle h\text{-}\rm SM\text{-}SM} =
{ h_1 c_\theta + h_2 s_\theta \over v} \left [ 2 m_W^2 W_\mu^+ W^{\mu -} + m_Z^2 Z_\mu Z^\mu - \sum_f m_f \bar f f \right] \,.
\ee
The remaining term $\Lcal_{h\text{-}h\text{-}h} $ represents the trilinear couplings among $h_1$ and $h_2$,
 \be
\Lcal_{h\text{-}h\text{-}h} = 
-{\kappa_{111} \over 6} v ~h_1^3
-{\kappa_{112} \over 2} v ~h_1^2 h_2 
-{\kappa_{221} \over 2} v ~h_2^2 h_1 
-{\kappa_{222} \over 6} v ~h_2^3 ~,
\ee
where
\besub
\bal
\kappa_{111} &= 
 {3 m_{h_1}^2 \over v^2 } \left ( c^3_\theta - s^3_\theta \frac{\tilde g v} { 2 m_A} \right )~,
\\ 
\kappa_{112} &= 
 {2 m_{h_1}^2 + m_{h_2}^2 \over v^2 } s_\theta c_\theta \left (c_\theta + s_\theta \frac{\tilde g v} { 2 m_A} \right )~,
\\ 
\kappa_{221} &= 
 {m_{h_1}^2 + 2 m_{h_2}^2 \over v^2 } s_\theta c_\theta \left (s_\theta - c_\theta \frac{\tilde g v} { 2 m_A} \right ) ~,
 \\
 \kappa_{222} &= 
 {3 m_{h_2}^2 \over v^2 } \left ( s^3_\theta + c^3_\theta \frac{\tilde g v} { 2 m_A} \right )~\,.
\eal
\eesub
Note that the quartic couplings $\lambda_H, \lambda_2, \lambda_{H22}$ do not explicitly appear in the above interaction terms since they are fixed in terms of $v,m_{h_1},m_{h_2},\sin \theta, \tilde g$ and $m_A$:
\begin{align}
\label{eq:higgs_to_gauge}
& \lambda_H=\frac{c^2_\theta m_{h_1}^2+s^2_\theta m_{h_2}^2}{v^2}~, \nonumber\\
& \lambda_2=\tilde{g}^2 \frac{s^2_\theta m_{h_1}^2+c^2_\theta m_{h_2}^2}{4 m_A^2}~, \nonumber\\
& \lambda_{H22}=\tilde{g} s_{2\theta} \frac{m_{h_2}^2-m_{h_1}^2}{4 v m_A} \,.
\end{align} 
The couplings in Eq.~(\ref{eq:simple_lagrangian}) therefore are a function of the 5 new physics parameters $m_\chi,m_A,m_{h_2},\tilde{g},\sin\theta$.
The hidden sector gauge coupling $\tilde{g}$ acts as an overall normalization parameter for the DM interactions.

In what follows, we analyze how the DM relic density is generated
as well as the constraints and prospects for Direct DM Detection.

\subsubsection{Relic density}

In conventional WIMP scenarios, the DM relic density is inversely proportional to the thermally averaged DM annihilation cross-section into SM fermions.
In the case of multicomponent DM, the situation is more involved since there are additional important processes such as conversion of one DM component into another. This complicates the analysis and
we therefore solve the system of Boltzmann equations numerically
 (cf.~\cite{Belanger:2011ww,Aoki:2012ub,Karam:2016rsz}).
\begin{figure}
\centering{
\includegraphics[scale=0.5]{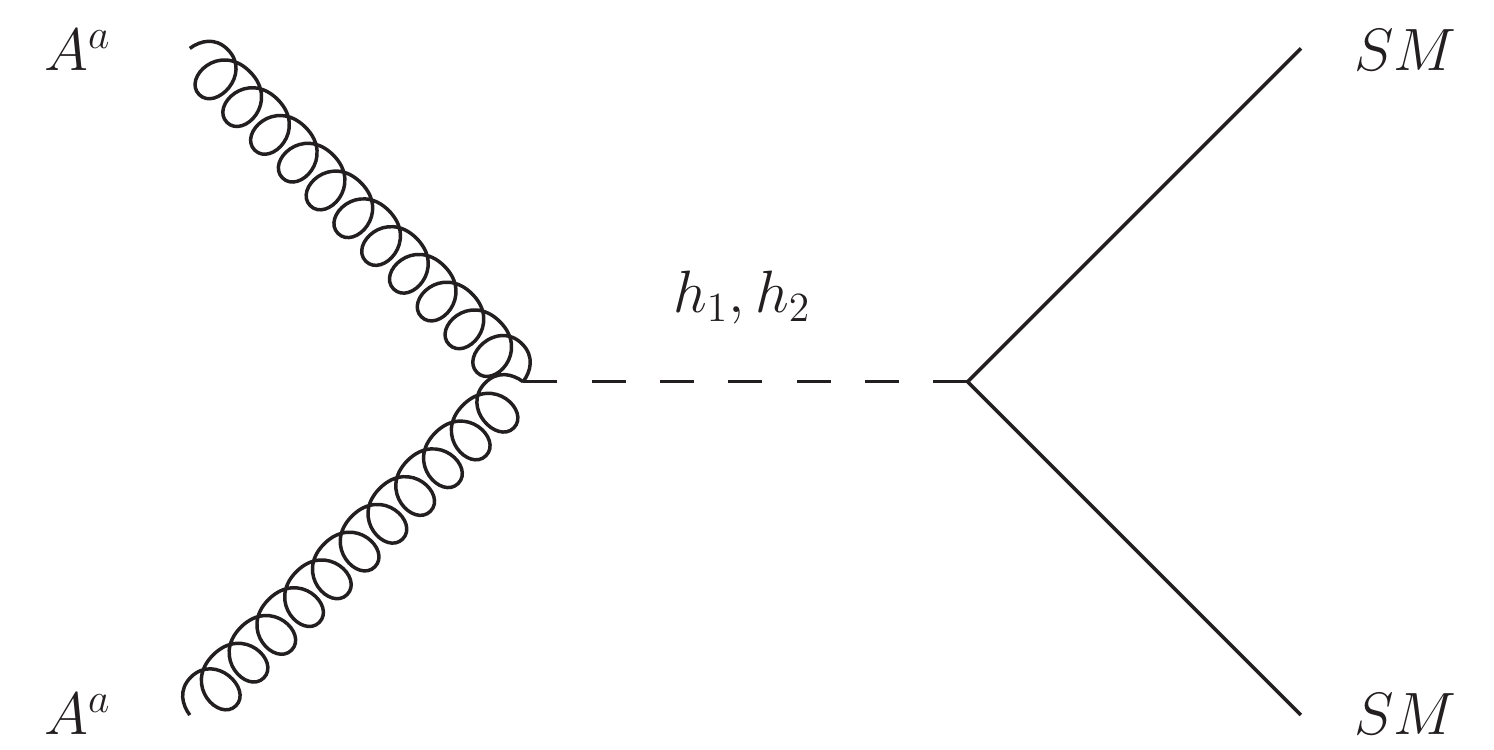}
\includegraphics[scale=0.5]{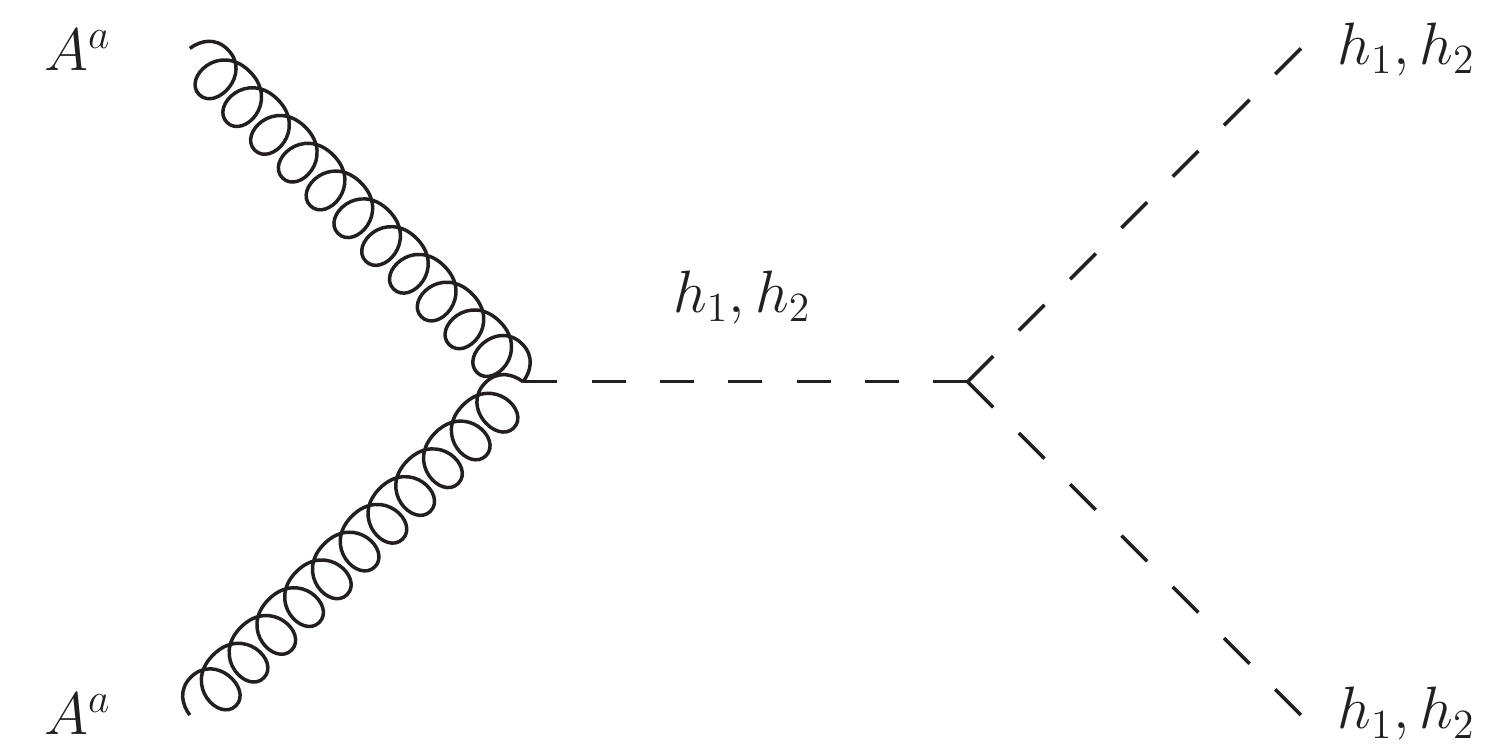}
\includegraphics[scale=0.5]{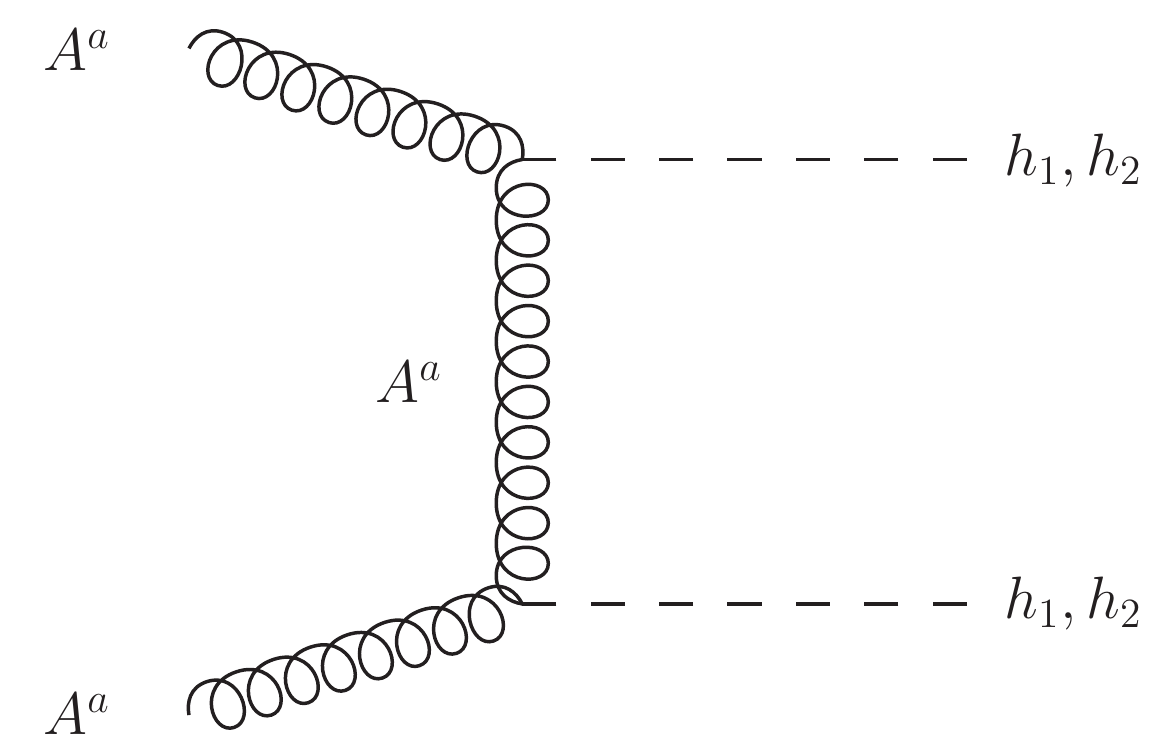}
\includegraphics[scale=0.5]{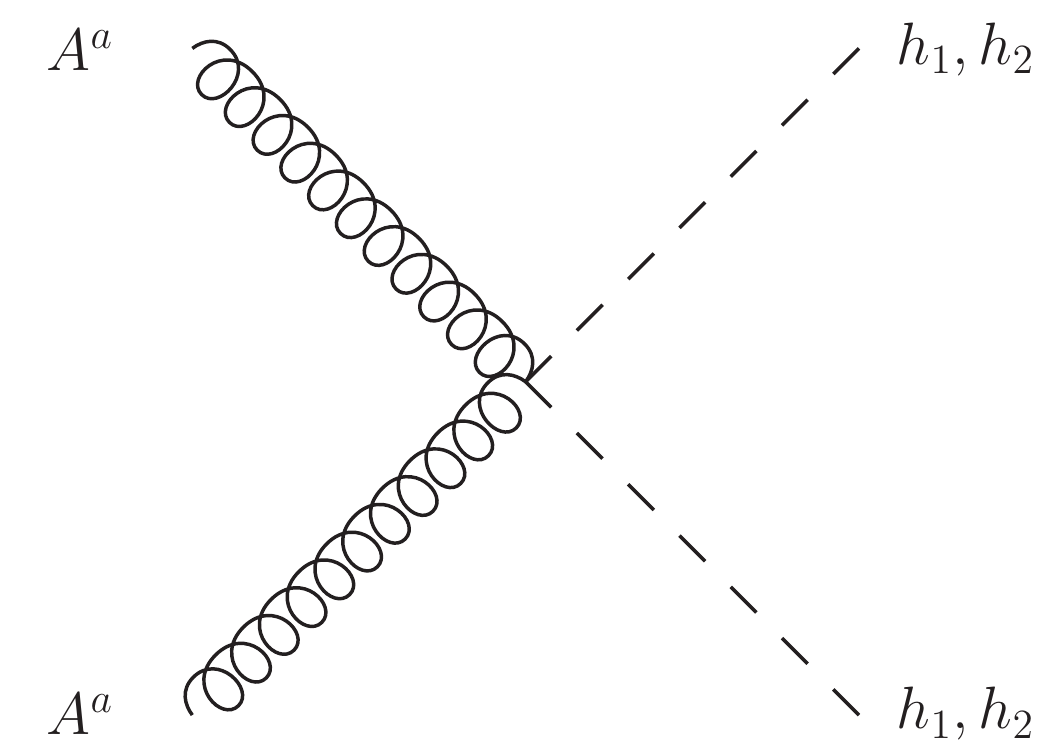}
 }
\caption{
Pair annihilation processes for the vectorial DM component.
}
\label{fig:pairV}
\end{figure}
\begin{figure}
\centering{
\includegraphics[scale=0.5]{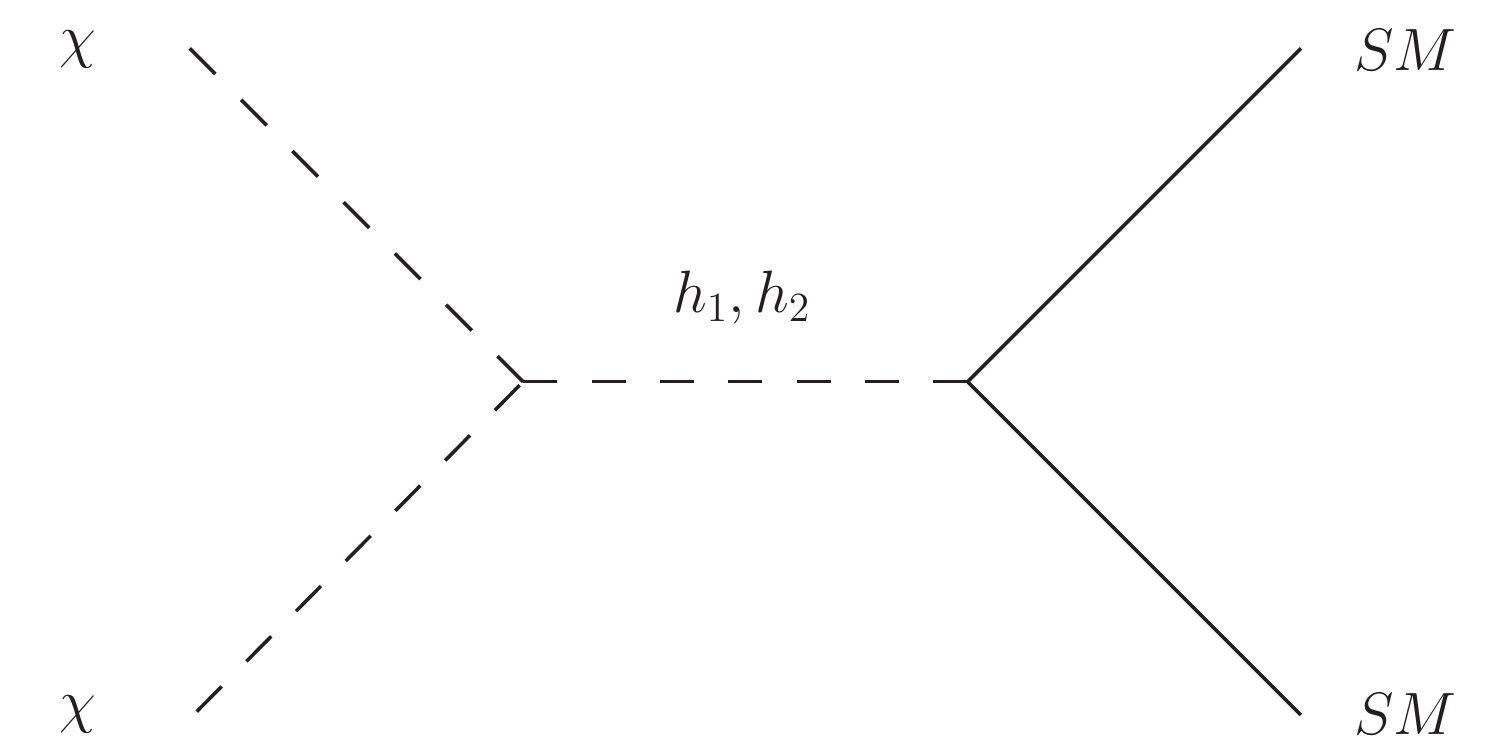}
\includegraphics[scale=0.5]{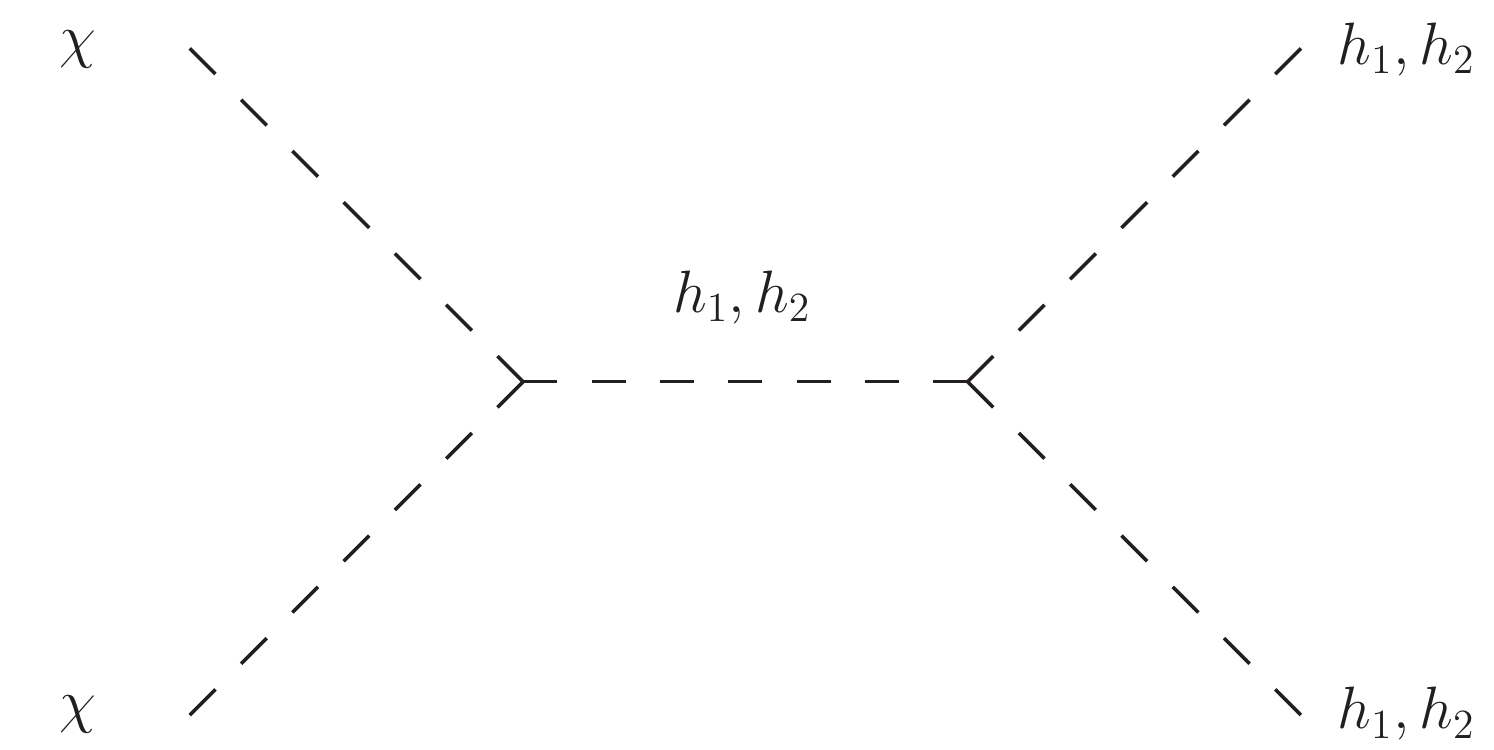}
 \includegraphics[scale=0.5]{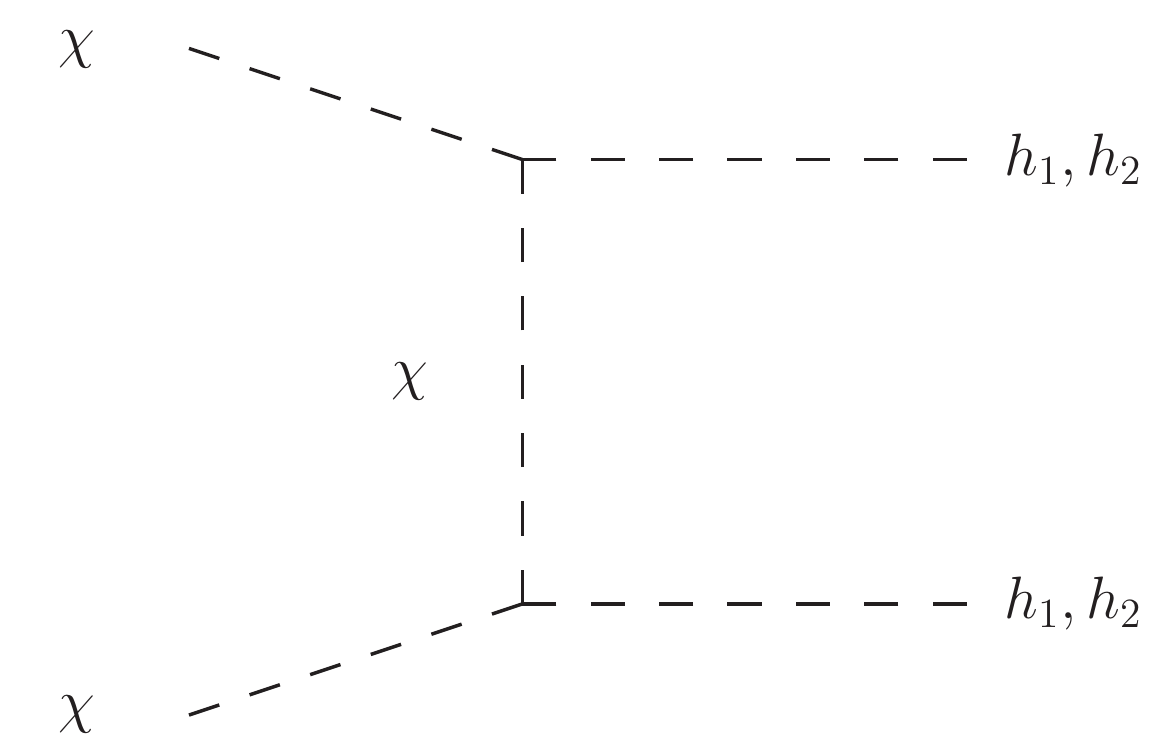}
\includegraphics[scale=0.5]{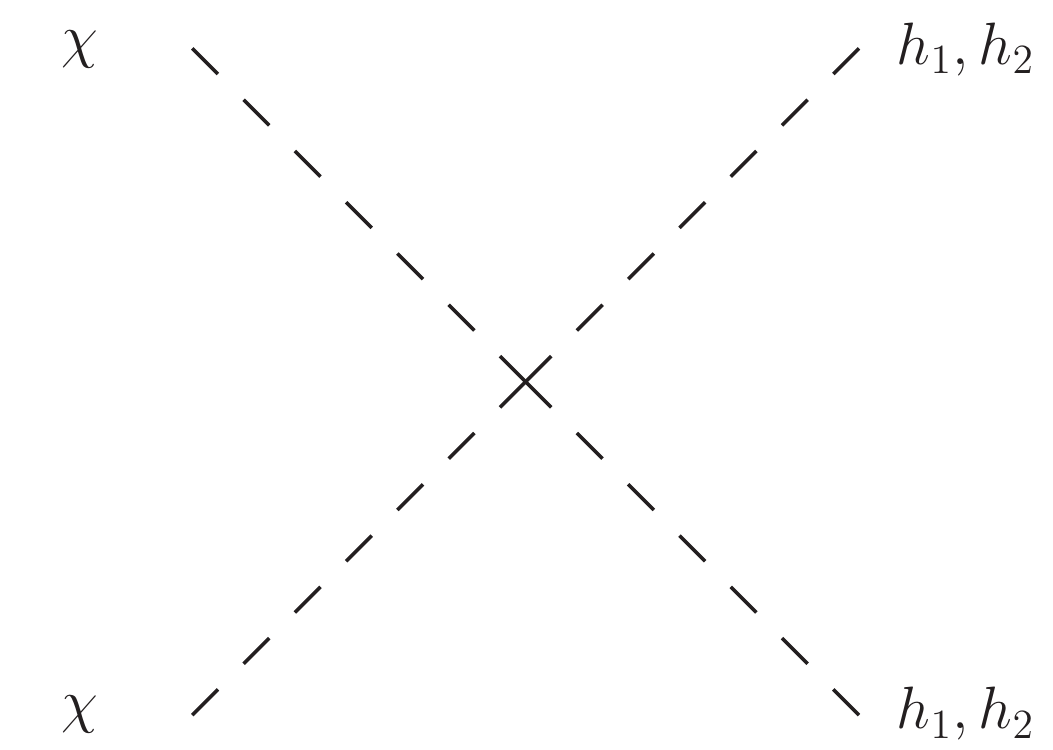}
 }
\caption{
Pair annihilation processes for the scalar DM component.
}
 \label{fig:pairS}
\end{figure}
\begin{figure}
\centering{
\includegraphics[scale=0.4]{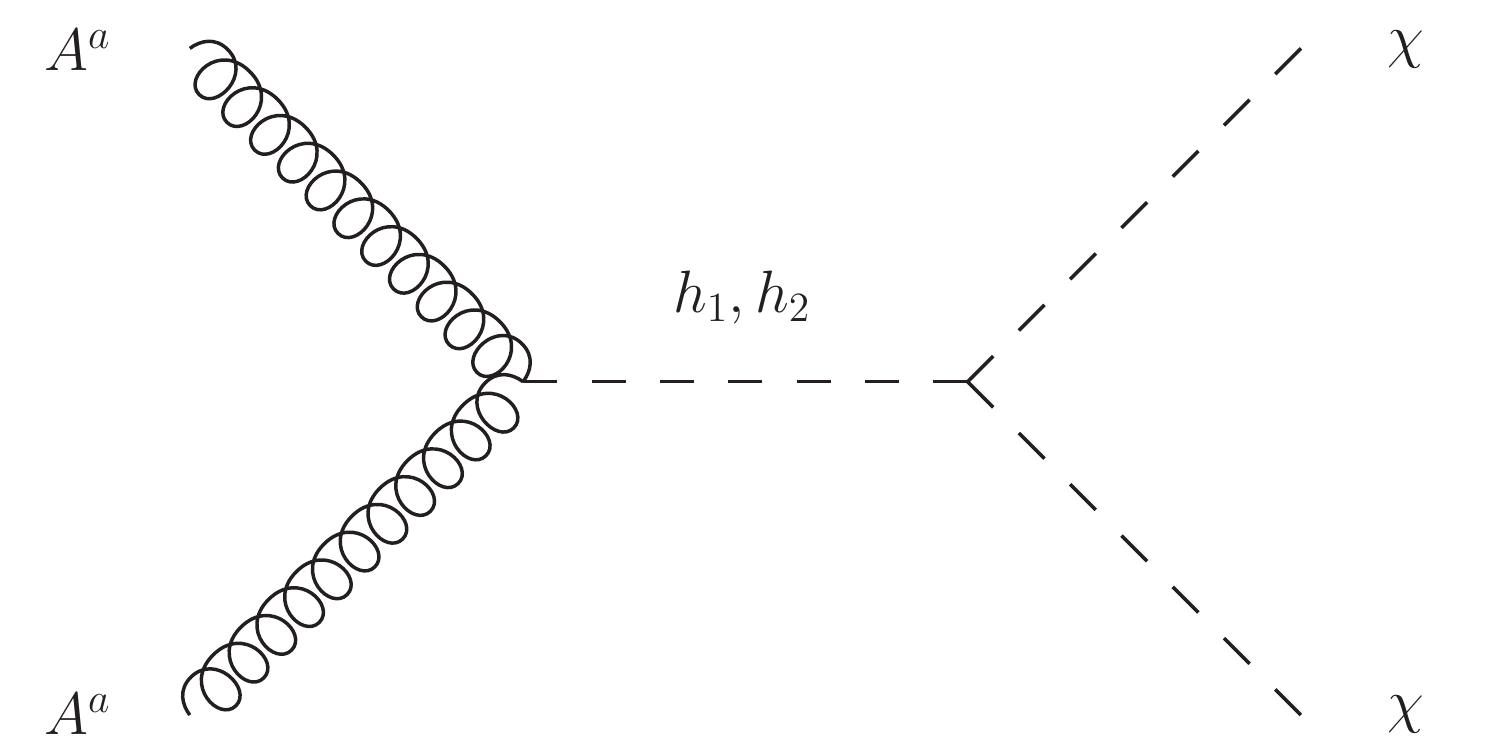}
 }
\caption{ 
Vector DM to scalar DM conversion.
}
\label{fig:conv}
\end{figure}
The Boltzmann equations are dictated by three types of processes:
\bi
\item
Pair annihilation of both DM components into SM fermions, gauge and Higgs bosons
\item
Conversion of one DM component into another: $AA \leftrightarrow \chi \chi$
\item
Semi--(co)annihilation (cf.~\cite{D'Eramo:2010ep,Belanger:2012vp}): $A A \rightarrow A_3 h_{1,2}$ and $A A_3 \rightarrow A h_{1,2}$ which changes the abundances of both the vector and the scalar component\footnote{$A_3$ decays to $\chi$ + SM matter.
We assume that this decay is fast enough so that we use 
 the Boltzmann equations with 2 DM components~\cite{Edsjo:1997bg}. If this is not the case, one must add an additional equation for the
 abundance of $A_3$ to Eq.~(\ref{eq:boltzmann_sys}).}
\ei
The relevant diagrams for the annihilation processes of the two DM components are presented in Figs.~\ref{fig:pairV}-\ref{fig:conv}, while the (subleading) semi--annihilation processes are not shown explicitly. 
The Boltzmann equations can be written as
\begin{align} \label{eq:boltzmann_sys}
 \frac{dY_A}{dx}&= -\overline{\langle \sigma v \rangle}_{AA \rightarrow XX} \left(Y_A^2-Y_{A,\rm eq}^2\right) - \overline{\langle \sigma v \rangle}_{AA \rightarrow \chi \chi} \left(Y_A^2-\frac{Y_{A,\rm eq}^2}{Y_{\chi,\rm eq}^{2}}Y_\chi^2\right)
\\
& \qquad -\overline{\langle \sigma v \rangle}_{AA \rightarrow A_3 h_{1,2}} \left(Y_A^2 - \frac{Y_\chi}{Y_{\chi,\rm eq}} Y^2_{A,\rm eq}\right),
\nonumber\\
\frac{dY_\chi}{dx} &= -\overline{\langle \sigma v \rangle}_{\chi \chi \rightarrow XX} \left(Y_\chi^2-Y_{\chi,\rm eq}^2\right)+\overline{\langle \sigma v \rangle}_{AA \rightarrow \chi \chi} \left(Y_A^2-\frac{Y_{A,\rm eq}^2}{Y_{\chi,\rm eq}^{2}}Y_\chi^2\right)
\nonumber\\
& \qquad-\overline{\langle \sigma v \rangle}_{A A_3 \rightarrow A h_{1,2}}Y_A Y_{A_3,\rm eq} \left(\frac{Y_\chi}{Y_{\chi,\rm eq}}-1\right)+ \overline{\langle \sigma v \rangle}_{AA \rightarrow A_3 h_{1,2}} \left(Y_A^2 - \frac{Y_\chi}{Y_{\chi, \rm eq}} Y^2_{A,\rm eq}\right), \nonumber
\end{align}
where $Y_i=n_i/s$ with $n_i$ being the corresponding number density and $s$ being the entropy, $x=m_A/T$ and
\begin{equation}
\overline{\langle \sigma v \rangle}(x)=\frac{\langle \sigma v \rangle s}{H x}\Big|_{T=m_A/x} \,,
\end{equation}
where $H$ is the Hubble rate.
\begin{figure}
\begin{center}
\includegraphics[width=7 cm]{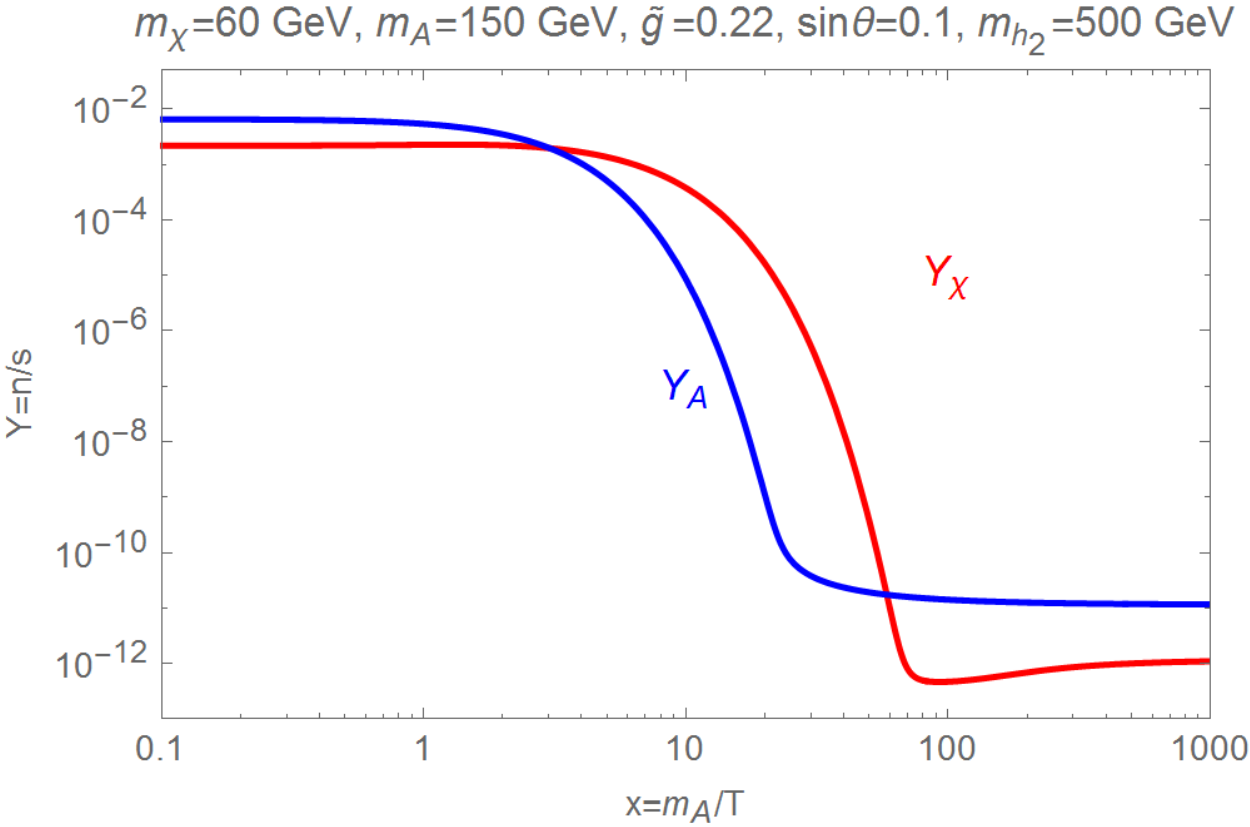}
\hspace{0.3cm}
\includegraphics[width=7 cm]{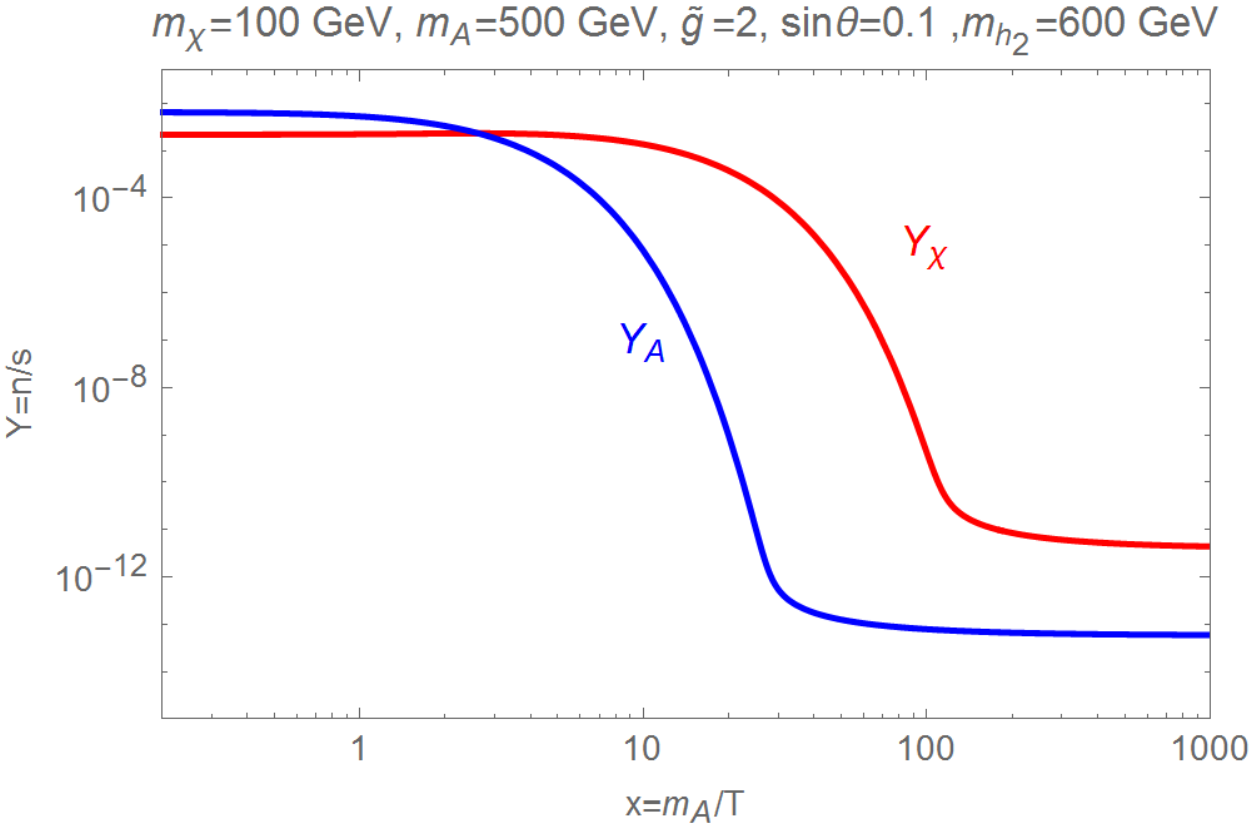}
\end{center}
\caption{Solutions of the Boltzmann equations for two benchmark parameter values. In each panel, the red and blue curves represent the abundances of the $\chi$ and $A$ DM components, respectively.}
\label{fig:relic_simplified}
\end{figure} 
The resulting evolution of the yields $Y_i$ for two benchmark parameter choices is shown in Fig.~\ref{fig:relic_simplified}. 
In the right panel, 
the relic density of the two components evolves similarly to that of conventional WIMPs,
i.e. it tracks the equilibrium distribution at Early times until decoupling. 
In the left panel, we see some modifications to this behaviour. In particular,
the pseudoscalar DM components annihilates very efficiently through the $h_1$ 
resonance which depletes its energy density, while at late times the $\chi$ fraction
of the DM number density increases due to the conversion process of Fig.~3.
In this case, the heavier DM component gives the dominant contribution to
the DM density.

Our numerical analysis (see below for more details) shows that the contribution of the processes $A A \rightarrow A_3 h_{1,2}$ and $A A_3 \rightarrow A h_{1,2}$ is negligible over most of the parameter space.
For a qualitative discussion of our numerical results, one may thus approximate the total DM relic density by the sum of the following two contributions~\cite{Gondolo:1990dk},
\begin{align}
\label{eq:relic_approximate}
 \Omega_{\rm DM,tot} h^2 
 &\approx 8.8 \times 10^{-11}\,{\mbox{GeV}}^{-2}\left[{\left(\bar{g}_{\rm eff,A}^{1/2}\int_{T_0}^{T_{f,A}} \!\! \langle \sigma v \rangle_{A} \frac{dT}{m_A}\right)}^{-1}
 \!\!\!\! +{\left(\bar{g}^{1/2}_{\rm eff,\chi}\int_{T_0}^{T_{f,\chi}} \!\! \langle \sigma v \rangle_{\chi} \frac{dT}{m_\chi}\right)}^{-1}\right]\nonumber\\ 
& \approx 8.8 \times 10^{-11}\,{\mbox{GeV}}^{-2}\left[\frac{x_{f,A}}{\bar{g}_{\rm eff,A}^{1/2}\left(a_A+x_{f,A}^{-1}b_A\right)}+\frac{x_{f,\chi}}{\bar{g}_{\rm eff,\chi}^{1/2}\left(a_\chi+x_{f,\chi}^{-1}b_\chi\right)}\right],
\end{align}
where $T_{f,\chi},T_{f,A}$ are the freeze-out temperatures of the two DM components, $T_0$ is the present time temperature and $g_{\rm eff,A,\chi}$ are the effective degrees of freedom in the Early Universe. 
In the second line of Eq.~(\ref{eq:relic_approximate}), we have used the velocity expansion $\langle \sigma v \rangle \simeq a+2b/x$ (using $\sigma v\simeq a+b v^2/3$ and $\langle v^2 \rangle=6/x$, cf. e.g.~\cite{Jungman:1995df}) and $x_{f,i}=m_{i}/T_{f,i}$.\footnote{This expansion is not valid in the vicinity of the s-channel poles. We note that
all results presented in this work rely on the full numerical calculation of the annihilation rates.}

In this work we only consider the case $m_\chi < m_A$ as required by our model.
Indeed, $A_3$ is always lighter than $A_{1,2}$ with our SU(3) breaking mechanism and $\chi$ must be lighter than $A_3$ to be stable. 
Hence, we include the conversion process $AA \rightarrow \chi\chi$, but not the reverse
(at least at late times).

The relevant annihilation cross-sections are s-wave dominated, i.e.~the coefficients $a_{\chi,A}$ are not suppressed. 
At leading order in velocity expansion, they read 
\begin{itemize}[leftmargin=*]
\item Pseudoscalar component:
\begin{align}
\label{eq:scalarxsection}
 \langle \sigma v \rangle_{\chi \chi \rightarrow \bar f f}
 =&\sum_f \frac{\tilde{g}^2 N_c^f}{4 \pi v^2 } s^2_{2 \theta} \ {\left(1-\frac{m_f^2}{m_\chi^2}\right)}^{3/2} 
 \frac{m_f^2 m_\chi^4 {\left(m_{h_1}^2- m_{h_2}^2\right)}^2}{{m_A^2 \left(m_{h_1}^2-4 m_\chi^2\right)}^2 {\left(m_{h_2}^2-4 m_\chi^2\right)}^2}~,
 \nonumber\\
 \langle \sigma v \rangle_{\chi \chi \rightarrow W^+ W^-}=
 &\frac{ \tilde{g}^2 }{2 \pi v^2 } s^2_{2 \theta} \sqrt{1-\frac{m_W^2}{m_\chi^2}} \left(1-\frac{m_W^2}{m_\chi^2}+\frac{3}{4}\frac{m_W^4}{m_\chi^4}\right) 
 \times \nonumber\\ & \quad 
 \frac{{m_\chi^6 \left(m_{h_1}^2- m_{h_2}^2\right)}^2}{{m_A^2 \left(m_{h_1}^2-4 m_\chi^2\right)}^2 {\left(m_{h_2}^2-4 m_\chi^2\right)}^2}~,
 \nonumber\\
 \langle \sigma v \rangle_{\chi \chi \rightarrow ZZ}
=&\frac{\tilde{g}^2 }{4 \pi v^2 } s^2_{2 \theta} \sqrt{1-\frac{m_Z^2}{m_\chi^2}} \left(1-\frac{m_Z^2}{m_\chi^2}+\frac{3}{4}\frac{m_Z^4}{m_\chi^4}\right) 
 \times \nonumber\\ & \quad 
\frac{{m_\chi^6 \left(m_{h_1}^2- m_{h_2}^2\right)}^2}{{m_A^2 \left(m_{h_1}^2-4 m_\chi^2\right)}^2 {\left(m_{h_2}^2-4 m_\chi^2\right)}^2}~.
\end{align}
\item Vector component: 
\begin{align}
\label{eq:vectorxsection}
\langle \sigma v \rangle_{AA \rightarrow \bar f f}=
&\sum_f \frac{\tilde{g}^2 N_c^f }{48 \pi v^2}s^2_{2 \theta} {\left(1-\frac{m_f^2}{m_A^2}\right)}^{3/2} \frac{{m_f^2 m_A^2 \left(m_{h_1}^2-m_{h_2}^2\right)}^2}{{\left(m_{h_1}^2-4 m_A^2\right)}^2 {\left(m_{h_2}^2-4 m_A^2\right)}^2}~,
\nonumber\\
\langle \sigma v \rangle_{AA \rightarrow W^+ W^-}=
& \frac{\tilde{g}^2 }{24 \pi v^2} s^2_{2 \theta} \sqrt{1-\frac{m_W^2}{m_A^2}} \left(1-\frac{m_W^2}{m_A^2}+\frac{3}{4}\frac{m_W^4}{m_A^4}\right) \times
\nonumber\\
& \frac{{m_A^4 \left(m_{h_1}^2-m_{h_2}^2\right)}^2}{{\left(m_{h_1}^2-4 m_A^2\right)}^2 {\left(m_{h_2}^2-4 m_A^2\right)}^2}
\nonumber\\
\langle \sigma v \rangle_{AA \rightarrow ZZ}= 
&\frac{\tilde{g}^2 }{48 \pi v^2}s^2_{2 \theta} \sqrt{1-\frac{m_Z^2}{m_A^2}} \left(1-\frac{m_Z^2}{m_A^2}+\frac{3}{4}\frac{m_Z^4}{m_A^4}\right) \times
\nonumber\\
&\frac{{m_A^4 \left(m_{h_1}^2-m_{h_2}^2\right)}^2}{{\left(m_{h_1}^2-4 m_A^2\right)}^2 {\left(m_{h_2}^2-4 m_A^2\right)}^2}~,
\nonumber\\
\langle \sigma v \rangle_{AA \rightarrow \chi \chi}=
&\frac{\tilde{g}^4}{768 \pi m_A^2}\sqrt{1-\frac{m_\chi^2}{m_A^2}} \times
\nonumber\\
&\frac{{\left(m_{h_1}^2 m_{h_2}^2-2 m_A^2 (m_{h_1}^2+m_{h_2}^2)+2 m_A^2 (m_{h_1}^2-m_{h_2}^2) c_{2\theta}\right)}^2}{{\left(m_{h_1}^2-4 m_A^2\right)}^2 {\left(m_{h_2}^2-4 m_A^2\right)}^2} \,.
\end{align}
\end{itemize}

The ``dark'' annihilation process $AA \rightarrow \chi \chi$ can be the most efficient
$A$--annihilation channel since it is not suppressed by $\sin^2 2 \theta$, which
is subject to rather tight experimental constraints \cite{Falkowski:2015iwa}. 
 This is because the process involves only the dark sector states. As a result, the annihilation cross-section of the vector DM component is often enhanced compared to that of the scalar component.

\begin{figure}
\begin{center}
\includegraphics[width=7 cm]{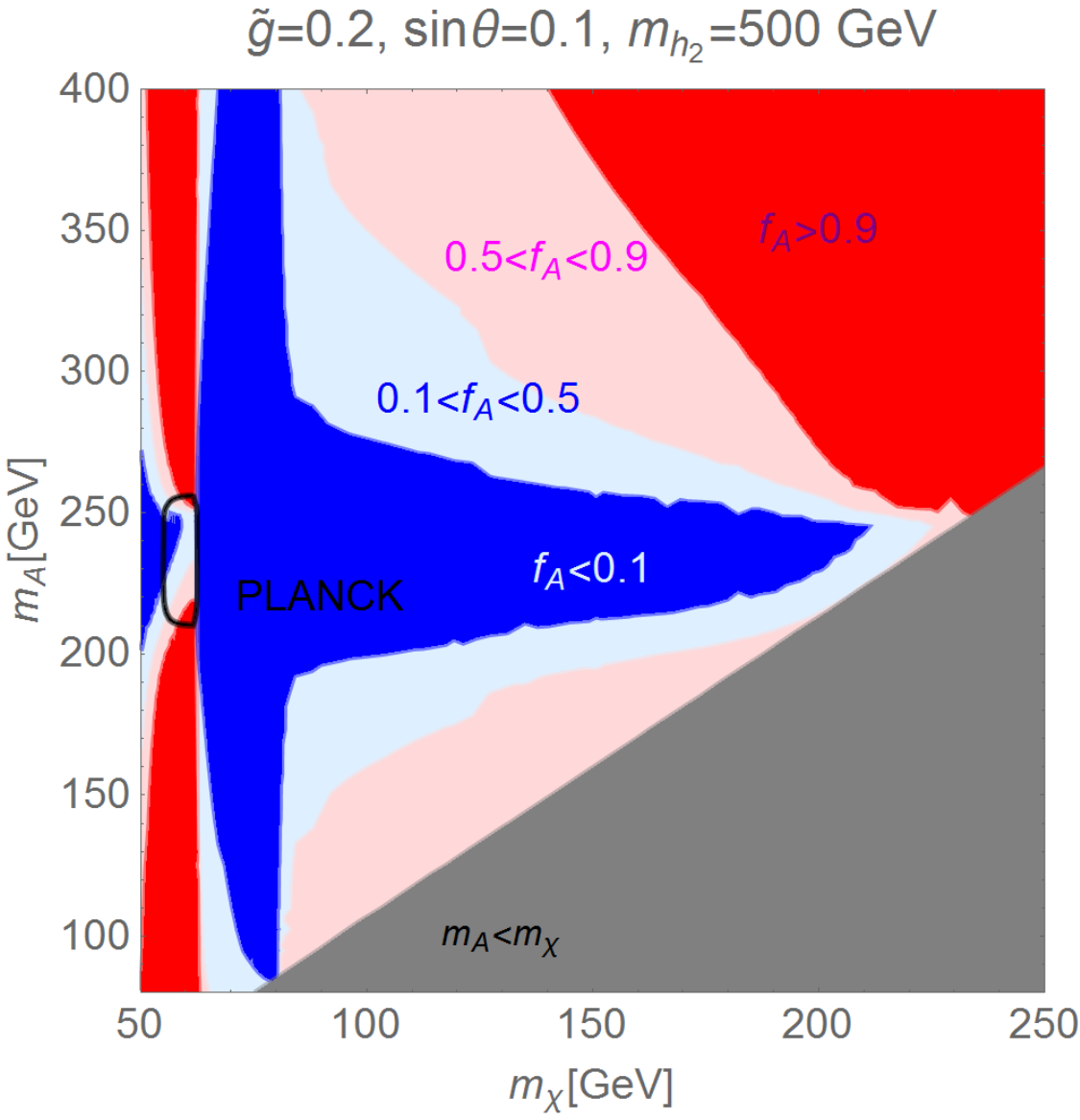}
\hspace{0.3cm}
\includegraphics[width=7 cm]{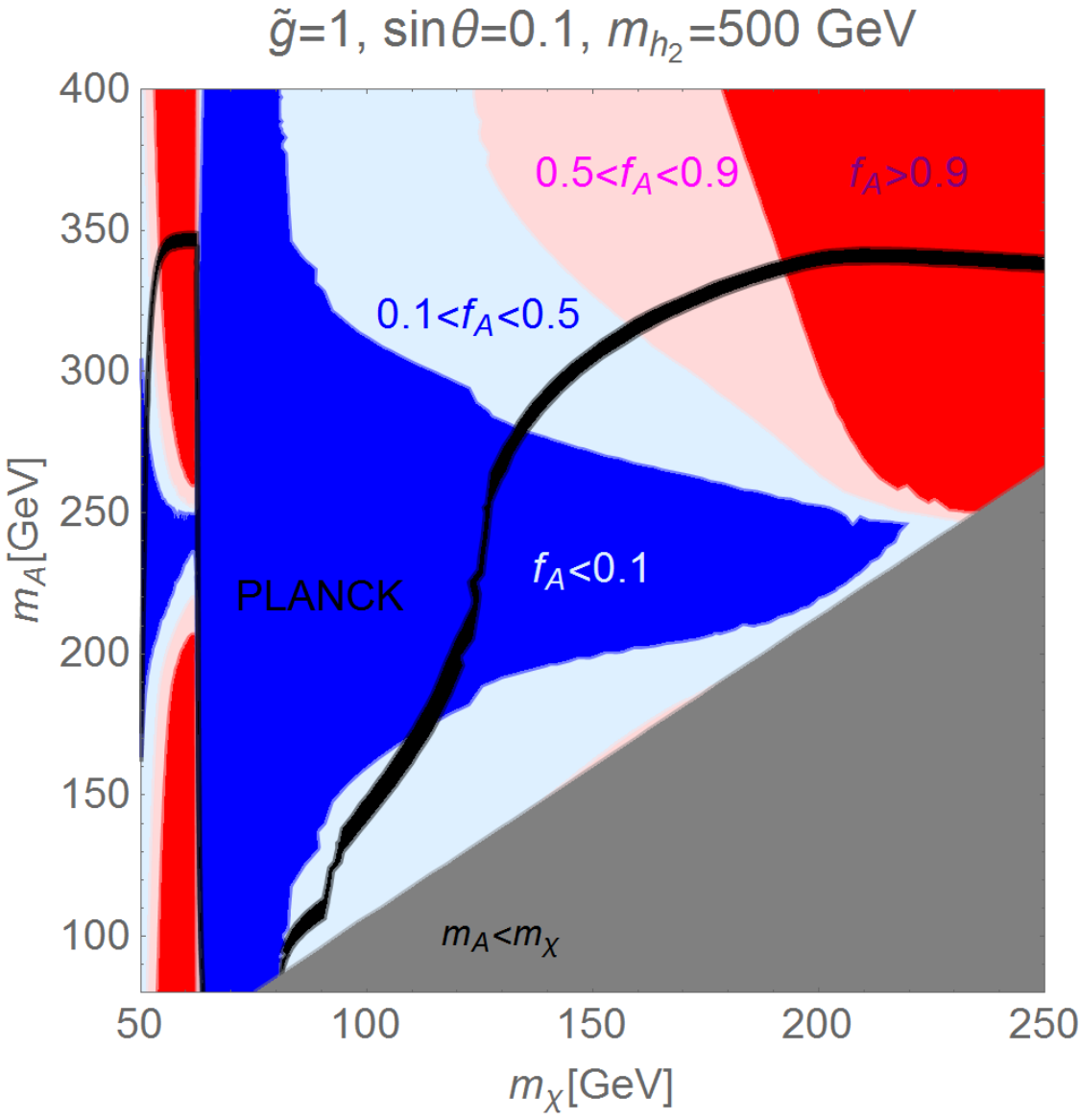}
\end{center}
\caption{The ratio $f_A=\Omega_A/ \Omega_{\rm tot}$ in the $(m_\chi,m_A)$-plane, for $\sin \theta = 0.1$, $m_{h_2}= 500~\GeV$ and $\tilde{g}=0.2$ (left) respectively $\tilde{g}=1$ (right). 
The blue, light blue, light red and red regions correspond to $f_A<0.1$, $0.1 < f_A < 0.5$, $0.5<f_A<0.9$ and $f_A>0.9$, respectively.
In the black regions, the observed total DM relic density is correctly reproduced at the 3 $\sigma$ level.
}
\label{fig:contours}
\end{figure} 
In Fig.~\ref{fig:contours}, we show the contribution of the vector component to the total DM relic density, $f_A=\Omega_A/ \Omega_{\rm DM,tot}$, in the plane $(m_\chi,m_A)$ with fixed $\tilde{g}$, $s_\theta$ and $m_{h_2}$. We distinguish the following three regions:
$f_A<0.1$ (blue), $0.1 < f_A < 0.5$ (light blue), $0.5<f_A<0.9$ (light red) and $f_A>0.9$ (red).
The correct DM relic density is only reproduced in the black regions,
so the purpose of the plot is to help understand how the composition of DM evolves as a function of parameters.

 Since the total DM relic density is given approximately by Eq.~(\ref{eq:relic_approximate}) and $x_{f,A} \approx x_{f,\chi}$, $\sqrt{\bar{g}_{\rm eff,A}}\approx \sqrt{\bar{g}_{\rm eff,\chi}}$, $f_A$ mostly depends on the ratio of the pair annihilation cross-sections of the two DM components:
\be
f_A\approx \frac{\frac{\langle \sigma v \rangle_{\chi}}{\langle \sigma v \rangle_{A}}}{1+\frac{\langle \sigma v \rangle_{\chi}}{\langle \sigma v \rangle_{A}}} \;.
\ee
 An obvious feature of Fig.~\ref{fig:contours}, which follows immediately from the above equation, is that the $A_{\mu}$ DM component dominates when $\chi$ annihilates resonantly, and vice versa.
For the regions away from the resonances, a closer inspection of $\frac{\langle \sigma v \rangle_{\chi}}{\langle \sigma v \rangle_{A}}$ is required.

Let us identify qualitative features of $\frac{\langle \sigma v \rangle_{\chi}}{\langle \sigma v \rangle_{A}}$.
For the mass range shown in the plot, $\langle \sigma v \rangle_{\chi}$ is dominated 
by $\langle \sigma v \rangle_{\chi \chi \rightarrow \bar b b}$ for $m_W > m_\chi$ and by $\langle \sigma v \rangle_{\chi \chi \rightarrow WW}$ for $m_W < m_\chi$.\footnote{There is also a sizeable contribution from the $\bar t t$ channel for $m_{\chi} > m_t$.}
 $\langle \sigma v \rangle_{A}$ has contributions from annihilation to both dark and visible sector final states.
Which one dominates depends mostly on the ratio $\tan \theta/ \tilde g$.
For instance, one has
\be \label{eq:vistodark}
\frac{\langle \sigma v \rangle_{AA \to W^+ W^-}}{\langle \sigma v \rangle_{A A \to \chi \chi}} = 8 \frac{\tan^2 \theta}{\tilde g^2}\frac{m^2_A}{v^2} \left(1-\frac{m_\chi^2}{m_A^2}\right)^{-1/2}
\times \left( 1+ \Ocal \left(\frac{m_W^2}{m_A^2},\frac{m_{h_1}^2}{4m_A^2},\frac{m_{h_1}^2}{m_{h_2}^2}\right)\right) .
\ee
The ratio $\frac{\langle \sigma v \rangle_{AA \to \bar b b}}{\langle \sigma v \rangle_{A A \to \chi \chi}}$ has an additional $m_b^2/m_\chi^2$ suppression factor.

\bi
\item
From Eq.~(\ref{eq:vistodark}) we see that in the right plot, where $\tilde g \gg \sin \theta$, the dark annihilation  $A A \to \chi \chi$  dominates in most mass regions. An exception is the region where $A_\mu$ is not much heavier than $\chi$ so that the dark annihilation is phase-space suppressed.

\bi
\item
For $m_\chi>m_W$, the  ratio $\frac{\langle \sigma v \rangle_{\chi}}{\langle \sigma v \rangle_{A}}$ becomes
\begin{align}
\frac{\langle \sigma v \rangle_{\chi \chi \rightarrow W^+ W^-}}{\langle \sigma v \rangle_{A A \rightarrow \chi \chi}} 
&= 
\\
&\!\!\!\!\!\!\!\!\!\! 96 \frac{\tan^2 \theta}{\tilde g^2}\frac{m^2_\chi}{v^2} \frac{ {\left(m_{h_2}^2-4 m_A^2\right)}^2}{ {\left(m_{h_2}^2-4 m_\chi^2\right)}^2}
\left( 1+ \Ocal \left(\frac{m_W^2}{m_\chi^2},\frac{m_{h_1}^2}{4m_\chi^2},\frac{m_{h_1}^2}{m_{h_2}^2},\frac{m_\chi^2}{m_A^2}\right)\right). \nonumber
\end{align}
For most parameter ranges of interest, 
 the $A_\mu$ annihilation cross section is much larger than that for $\chi$. As a result, $f_A<0.1$.
This does not however apply to the region $m_A^2 \gg m_{h_2}^2/4$ (upper part of the plot) in which case the factor $(2 m_A/m_{h_2})^4$ can compensate the small ratio $(\tan \theta/\tilde g)^2$.
\item
For $m_\chi<m_W$, the $\chi $ annihilation cross section is suppressed by the $b$--quark mass. The resulting  $\frac{\langle \sigma v \rangle_{\chi}}{\langle \sigma v \rangle_{A}}$ and 
$f_A$ are small unless 
$\chi $ annihilates resonantly.
\ei

\item
In the left plot, $\sin \theta$ and $\tilde g$ have similar sizes so that the visible and dark $A_\mu$ annihilation channels play in general
comparable roles. Two features are clearly visible:
 resonant $A_\mu$ annihilation
or $b$--quark mass suppression of $\langle \sigma v \rangle_{\chi}$ for $m_W > m_\chi$
lead to small $f_A$.
\ei

We find that the correct relic DM density typically requires sizable $\tilde{g}$ and $s_\theta$.
If $\tilde{g}$ is too small, the $\chi$ DM component is overproduced.
As seen in Fig.~\ref{fig:contours}, at $\tilde{g} =0.2$ one must
resort to resonant $\chi$ annihilation to keep its density under control. The DM composition is very sensitive to the exact $\chi$--mass
 in this case. With a larger gauge coupling, $\tilde{g}=1$,
 the correct relic density is achieved in substantial regions of parameter space. We find numerically that both DM components can be as heavy as a few hundred GeV, while $ \tilde g s_\theta \gtrsim 0.01$ is required to 
 keep the $\chi$--annihilation efficient. While qualitative features 
 of the plot can be understood semi-analytically, we have performed our numerical analysis using the software Micromegas~\cite{Belanger:2014vza} which is well suited for 2 component DM.

\begin{figure}[t]
\begin{center}
\includegraphics[width=6.8 cm]{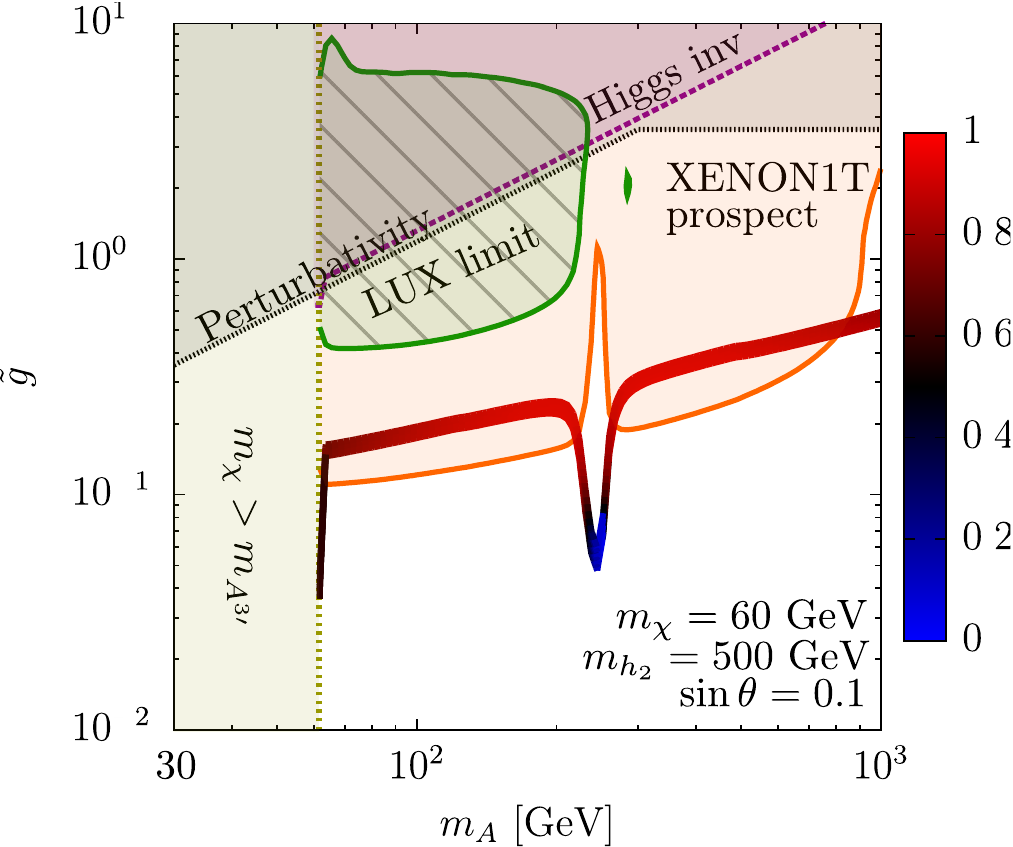}
\hspace{0.3cm}
\includegraphics[width=6.8 cm]{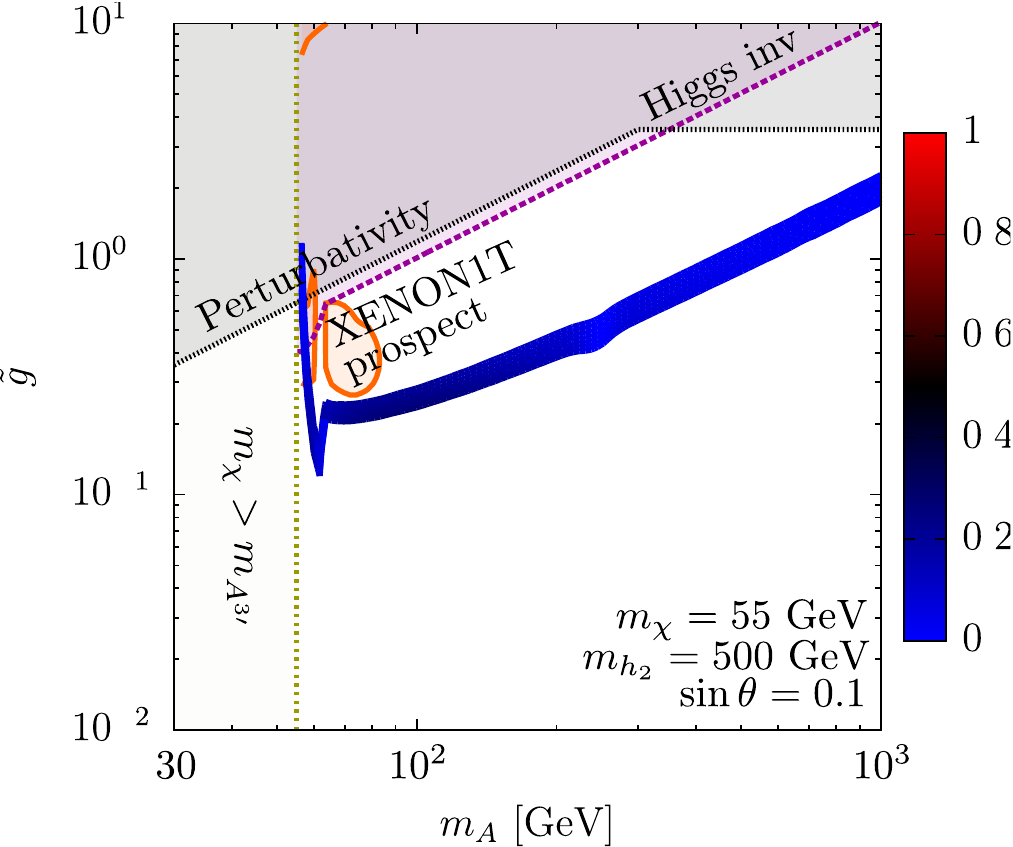}\\
\includegraphics[width=6.8 cm]{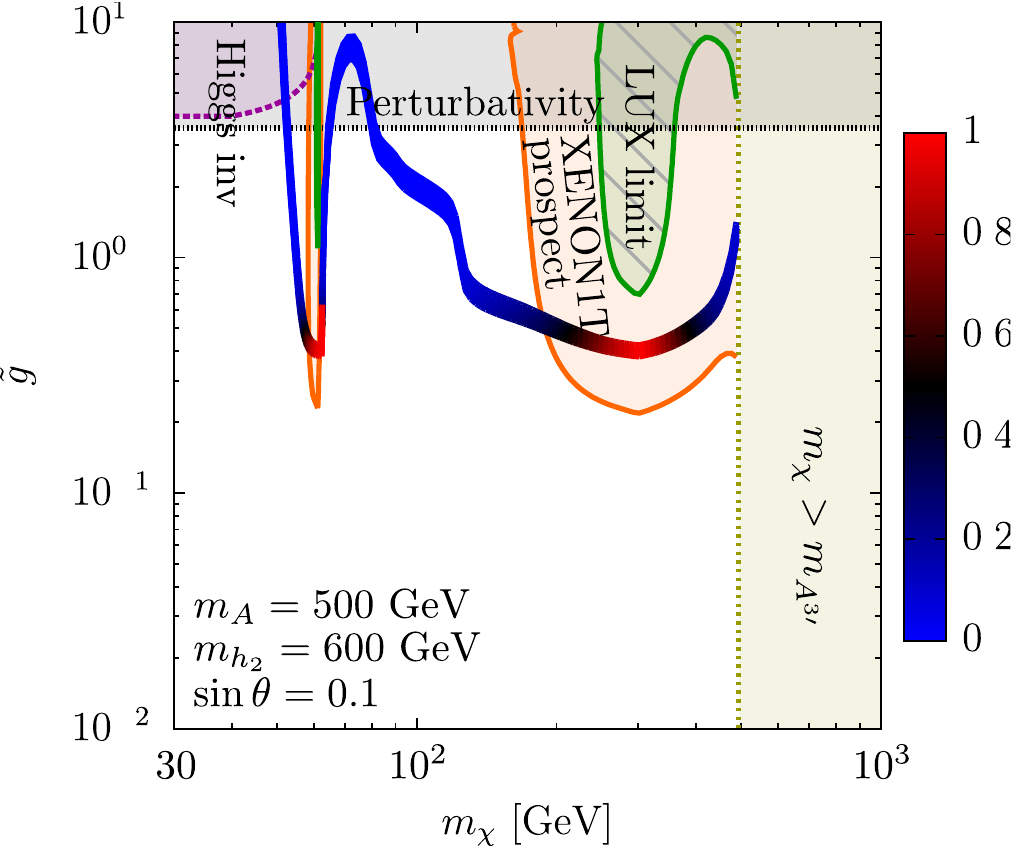}
\hspace{0.3cm}
\includegraphics[width=6.8 cm]{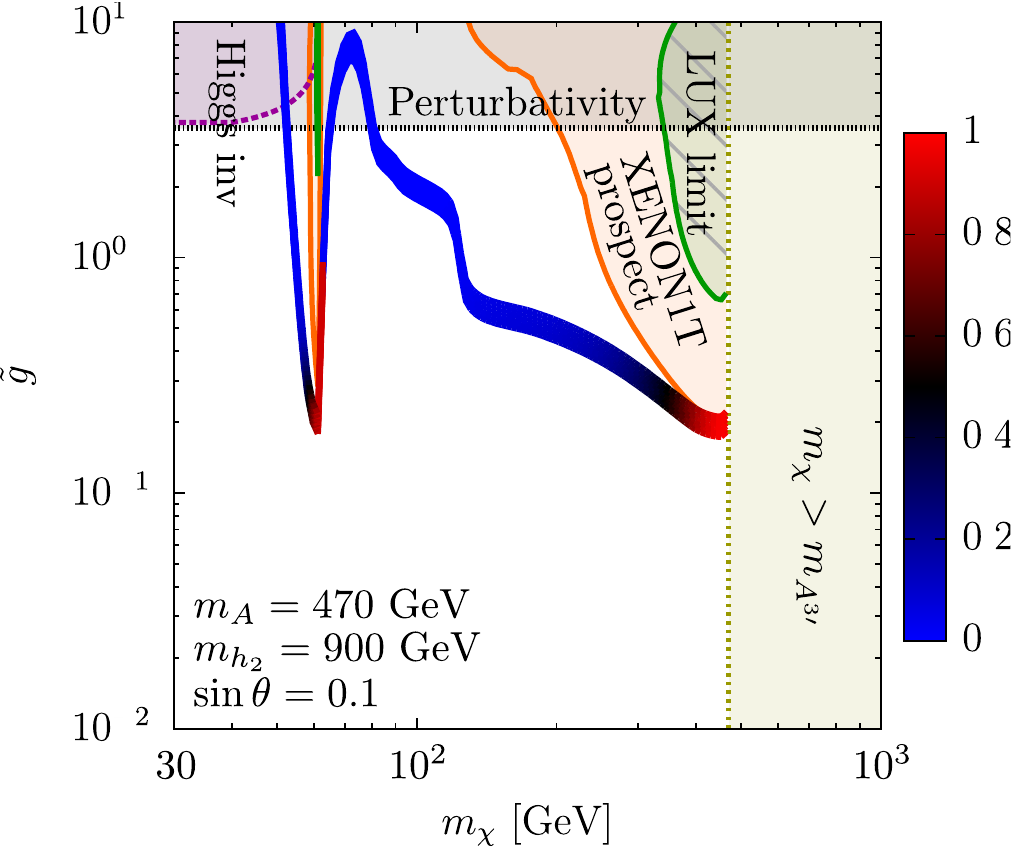}
\end{center}
\caption{ Dark matter constraints
 in the plane $(m_A,\tilde{g})$ (upper panel) and $(m_\chi,\tilde{g})$ (lower panel) for $v_1 \gg v_2$. 
The blue--red band indicates the correct relic DM density with the 
blue (red) end of the spectrum referring to the spin-0 (spin-1) component dominance. The other curves mark the following constraints:
grey -- perturbativity, purple -- invisible Higgs decay, green -- LUX 2016 direct DM detection, orange -- XENON1T direct DM detection prospects. 
 }
\label{fig:dm1}
\end{figure}

In Fig.~\ref{fig:dm1}, we show the contours of correct DM relic density in the $(m_A,\tilde{g})$-plane (left panels) and $(m_\chi,\tilde{g})$-plane (right panels). 
The color coding along the contours indicates the value of $f_A$:
the red (blue) end of the spectrum refers to vector (pseudoscalar) dominance.

Many features of the plots can be understood qualitatively.
In the upper left panel, the dark annihilation process $AA \rightarrow \chi\chi$ is important, yet the resulting $\chi$ states annihilate very efficiently through the $h_1$ 
resonance into the SM fields. As a result, DM is mostly vector (apart from the small region $m_A \simeq m_{h_2}/2$). The necessary $\tilde g$
at $\sin\theta=0.1$ is smaller than that in \cite{Gross:2015cwa} due to the availability
of the dark annihilation channel, albeit it remains in the same ballpark of ${\cal O}(10^{-1})$. In the right upper panel, the $\chi$ mass moves a bit further from the center of the $h_1$ resonance, which changes the DM composition and requires somewhat larger gauge couplings. Nevertheless, the resonance is still efficient and allows one to obtain 
the correct relic density with a relatively small $\tilde g$.

In the lower panels, the relic density band has the resonant structure similar to that of 
\cite{Gross:2015cwa}. The narrow $h_1$ resonance at $m_\chi \simeq m_{h_1}/2$
 is followed by a much broader\footnote{The reason is the large width of $h_2$ due to many available decay channels as well as the thermal averaging effect which makes DM annihilation efficient even away from $m_{h_2}/2$.} resonance around $m_{h_2}/2$. The kinks in the band represent new annihilation channels becoming kinematically available, e.g. $\chi\chi
 \rightarrow h_1 h_1$. In most regions away from the tip of the resonance, DM is predominantly
 pseudoscalar.

Besides the prospects for direct detection, which will be discussed in the following subsection, Fig.~\ref{fig:dm1} displays the limits from perturbativity of the quartic
($\lambda_i< 4\pi$) and gauge ($\tilde g^2_i< 4\pi$)
 couplings as well as those from the invisible decay of the 125 GeV Higgs boson.
 While the former has almost no impact on the region with the correct DM relic density, the latter excludes light values of $m_\chi$ below approximately 50 GeV.

\subsubsection{Direct detection}

In this subsection, we discuss the limits from the LUX experiment~\cite{Akerib:2015rjg}, as well as prospects for direct detection in XENON1T~\cite{Aprile:2015uzo}. 
The interactions of the pseudoscalar and vector DM components with nuclei are vastly different, thus it is convenient to discuss them separately.

\bi
\item Scattering of the $\chi$ component:\\ \ \\
The spin--independent (SI) $\chi$--nucleon scattering cross-section vanishes at tree-level in the limit of low momentum transfer:
\be
\sigma_{\chi N}\simeq 0 \,.
\ee
Thus, the pseudoscalar DM component appears to hide from detection
albeit in a different manner compared to the known mechanisms which rely
on the pseudo-scalar/axial-vector mediators~\cite{Boehm:2014hva,Lebedev:2014bba}. 
The reason for it is a cancellation between the $t$--channel $h_1$ and $h_2$ contributions
which follows from the coupling 
\be
\Lcal \supset (1+r) \frac{\tilde{g}}{2 m_A} \left(-h_1 m_{h_1}^2 s_\theta +h_2 m_{h_2}^2 c_\theta\right) \chi^2 
\ee
as well as the $h_1,h_2$ couplings to SM matter.

Since this is an important feature of this model, let us discuss the origin of this `blind' spot in the $\chi$--$N$ scattering in more detail.
To this end, let us consider now the $\chi$--nucleon interaction the interaction basis, i.e.~before diagonalising the scalar mass matrix.
The pseudoscalar $\chi$ interacts with the scalars $\varphi_1$ and $\varphi_2$ of the 
dark sector and $h$ of the Standard Model, while only the latter couples to quarks.
In the interaction basis, the effective $\chi \chi NN$ coupling is 
\begin{equation}
g_{\chi \chi NN}= (\vec{\kappa}_\chi)^{\dagger} ({\bold{m}}^{2})^{-1} \vec{\kappa}_f \;,
\end{equation}
with 
\begin{equation}
\label{eq:m1}
\vec{\kappa}_\chi 
 \propto \left(
\begin{array}{c}
v \, \lambda_{H22} \\
v_1 (\lambda_3+\lambda_4-\lambda_5)\\
v_2 \, \lambda_2
\end{array}
\right),
\quad
\vec{\kappa}_f
 \propto
\left(\begin{array}{c}
k \\
0 \\
0
\end{array}
\right).
\end{equation}
Here $\vec{\kappa}_\chi$ represents the $\chi$ couplings to $h$,$\varphi_1$ and $\varphi_2$; $\vec{\kappa}_f$ gives the fermion couplings of $h$,$\varphi_1$ and $\varphi_2$; ${\bold{m}}^2$ is the upper left 3$\times$3 block of the {\it CP}-even state mass matrix given in Eq.~(\ref{massmat}).
One now easily finds that
\begin{equation}
\label{eq:fv}
 g_{\chi \chi NN} \propto \lambda_{H11} \lambda_2 - \lambda_{H22} \lambda_3 \,.
\end{equation}
We see that the reason for $\sigma_{\chi N} \simeq 0$ is that 
we have taken $\lambda_{H11},\lambda_3$ to be negligible.
 In other words, we have assumed that only one scalar mixing is significant, that is,
 between $h$ and $\varphi_2$, while the $\varphi_1-\varphi_2$ and 
 $h-\varphi_1$ ones are very small. Although this is just a simplifying assumption,
it is meaningful as one does not expect all the couplings to be equally significant.
The corresponding region of parameter space represents an ``alignment limit'' where
the $3\times 3$ mass matrix turns effectively into a $2\times2$ one. 
This yields a simple calculable model, which
could perhaps be justified in a framework of a more sophisticated UV completion.
Were we to relax our assumption, we would get contributions which are suppressed by the 
 $\varphi_1-\varphi_2$ and $h-\varphi_1$ mixing angles.

\item Scattering of the $A_\mu$ component:\\ \ \\
The $t$--channel exchange of $h_1,h_2$ leads to the following SI scattering cross-section on nuclei:
\begin{equation}
\label{eq:VSI}
\sigma_{A N}=\frac{{\tilde{g}}^2 \mu^2_{\rm A N}}{4\pi}s^2_\theta c^2_\theta {\left(\frac{1}{m_{h_1}^2}-\frac{1}{m_{h_2}^2}\right)}^2 \Bigl(f_p Z/A + f_n (1-Z/A)\Bigr)^2,
\end{equation}
where $\mu_{\rm A N}=m_A m_N/(m_A + m_N)$ and
\begin{equation}
f_N=m_N \left( \sum_{q=u,d,s} f^N_{T_q} \frac{y_q}{m_q}+\frac{2}{27}f^N_{T_G}\sum_{q=c,b,t}\frac{y_q}{m_q} \right), \quad \textrm{where}\quad N=n,p
\end{equation}
parametrizes the Higgs-nucleon coupling. 
(For an up-to-date determination of $f_N$ see e.g.~\cite{Crivellin:2013ipa}.)
In the above expression, $y_q$ are the SM Yukawa couplings, 
 $f^N_{T_q}$ denotes the contribution of quark $q$ to the mass of the nucleon $N$ and $f^N_{T_G}=1-\sum_{q=u,d,s} f^N_{T_q}$.
\ei

Since the SI scattering of $\chi$ on nuclei is suppressed, the Direct Detection limits are obtained by comparing the experimental limits to the rescaled cross-section $f_A \sigma_{NA}$.
The current limits from the LUX experiment and the projected sensitivity of XENON1T are shown in Fig.~\ref{fig:dm1} by green and orange contours, respectively, assuming the exposure time considered in~\cite{Akerib:2015rjg}. 
As discussed in the previous subsection, the $\chi$ component typically dominates the relic DM density which renders the current and future Direct Detection constraints weak to irrelevant.

Furthermore, even the regions dominated by the vector DM component
are hard to probe unless one employs 1 ton detectors and a few years of exposure. This is in contrast to ``typical'' Higgs portal DM models 
(see e.g.~\cite{Gross:2015cwa}). One reason for the difference is that our setup allows for dark annihilation $AA \rightarrow \chi \chi$ which
can be dominant. The presence of this additional channel lowers the
gauge coupling $\tilde g$ 
required by the correct relic abundance thereby diminishing the relevance of Direct DM Detection. In addition, a low value of $\sin\theta$ provides another suppression factor compared to the analysis of 
\cite{Gross:2015cwa}. 

Finally, let us note that the unusual shape of the LUX/XENON constraints in Fig.~\ref{fig:dm1} is due to the non--trivial composition of dark matter. For instance, keeping $m_A$ and $m_\chi$ 
fixed while increasing $\tilde g$ changes the DM composition factor $f_A$. At large enough $\tilde g$, the dark annihilation channel typically dominates which makes DM mostly pseudoscalar and thus not prone to Direct Detection. This 
feature is clearly visible in the plots.

%=========================================================================
\subsection{Case $v_1 \simeq v_2$} \label{sec:v1=v2}
%=========================================================================

In this subsection,
 we repeat our analysis for $v_1\simeq v_2$. More specifically, we take $v_1 = 1.2 \, v_2$ in our numerical studies.
The main difference from the previous case is that all hidden gauge bosons have comparable masses now, cf. Eq.~(\ref{eq:vmasses_1}).
Also the scalars ${\cal H},{\tilde \varphi_3}$ are expected to be as heavy as $h_1,h_2$. However, we will focus on the parameter region where ${\cal H},{\tilde \varphi_3}$ are heavier than the other scalars and their effect can be neglected for our purposes. This is a simplifying assumption
which makes our numerical analysis tractable.

\subsubsection{Relic density}

\begin{figure}[t]
\begin{center}
\includegraphics[width=6.8 cm]{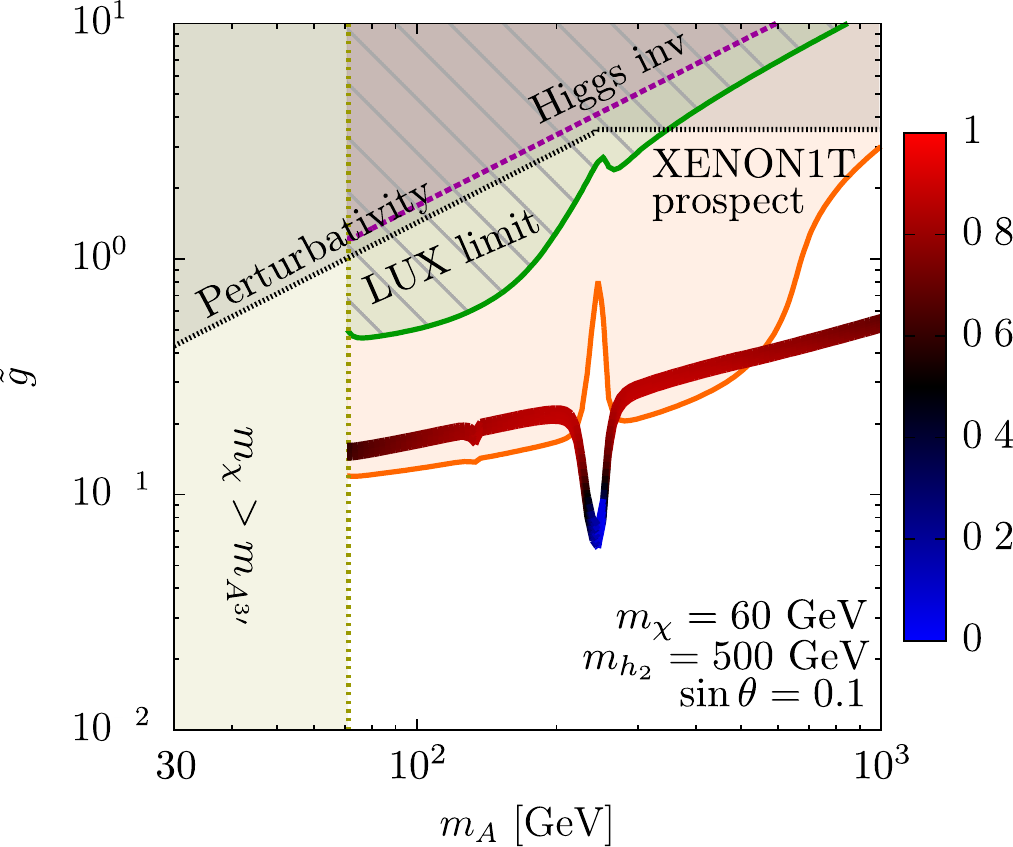}
\hspace{0.3cm}
\includegraphics[width=6.8 cm]{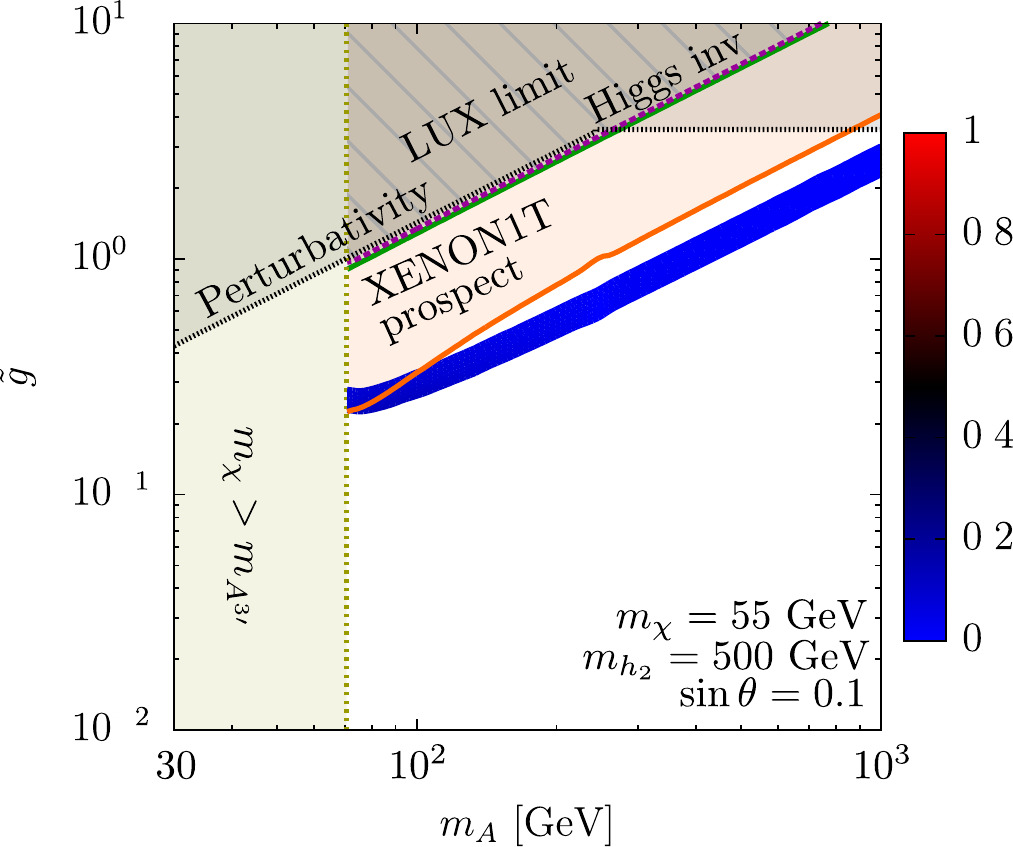}\\
\includegraphics[width=6.8 cm]{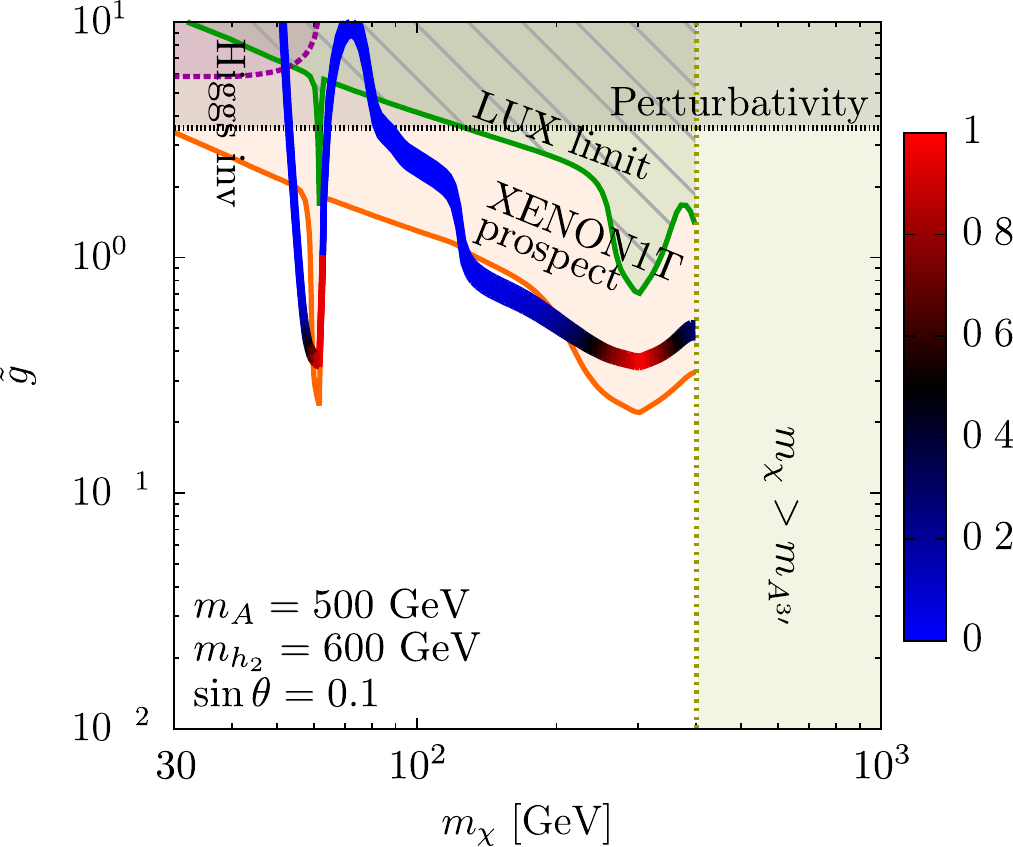}
\hspace{0.3cm}
\includegraphics[width=6.8 cm]{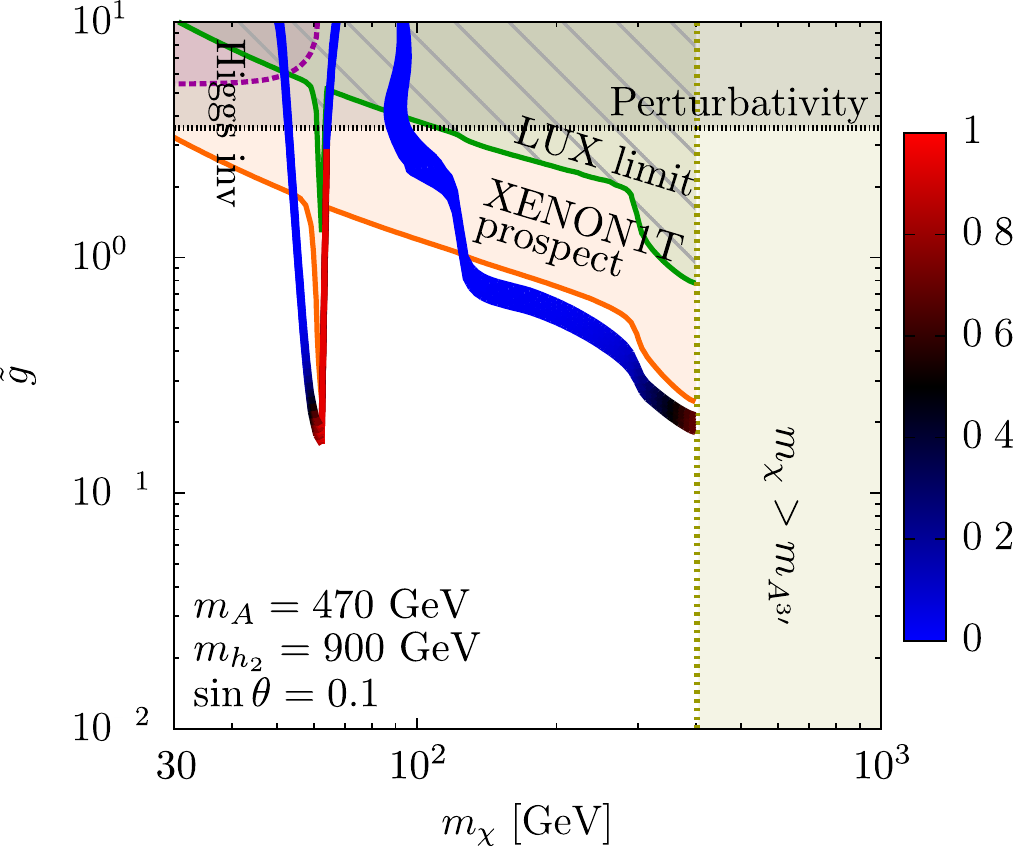}
\end{center}
\caption{Dark matter constraints as in Fig.~\ref{fig:dm1} for $v_1 =
1.2 \, v_2.$}
\label{fig:dm2}
\end{figure}

Although the general structure of the Boltzmann equations~(\ref{eq:boltzmann_sys}) is not altered, the larger number of processes makes a semi--analytic treatment very complicated in the case $v_1 \sim v_2$. 
Therefore we only perform the full numerical analysis with Micromegas. 
Compared to the $v_1 \gg v_2$ case, the following additional processes occur:
\bi
\item
The gauge bosons $A^{4-7},A'^8$, which are not decoupled now, act as additional mediators of annihilation processes and therefore can enhance the annihilation rates of the vector DM component.\footnote{
For $v_1/v_2$ very close to 1, coannihilation processes involving for example $A^4$ and $A^1$ play a role. For $v_1/v_2=1.2$, such processes are unimportant since they are typically suppressed by
$
\exp\left(-x_f (m_{A^4}-m_A)/m_A\right)\simeq 0.02. \nn
$
}
\item
Kinetic mixing terms give rise to additional interactions which 
scale approximately as $m_\chi^2/m_A^2$. 
Their impact is thus limited unless the two DM components have similar masses. 
\item
Self--interaction of the $A^i$ states could a priori lead to a sizeable effect. 
Our numerical analysis shows, however, that this is not the case.
\ei
In Fig.~\ref{fig:dm2}, we show the regions of correct DM relic density, for the same sets of parameters as in Fig.~\ref{fig:dm1} (apart from $v_1 = 1.2 \, v_2$). 
We see that the isocontours of correct relic abundance 
do not differ substantially from those for the case 
 $v_1 \gg v_2$.

\subsubsection{Direct detection}

As seen from Fig.~\ref{fig:dm2}, the Direct Detection limits change substantially.
Even though there is a cancellation in $\sigma_{\chi N}$ as described before, for
$v_1 \sim v_2$ it is incomplete. The mixing term $A^{\mu\,6}\partial_\mu \chi$ is now
important since $A^{\mu\,6}$ does not decouple. Eliminating this term by field redefinition
leads to an additional coupling that scales as $m_{A^6}^{-2}$. The resulting $\chi$--$N$ scattering cross section is then 
\be
\frac{\sigma_{\chi N}}{\sigma_{AN}} \simeq \frac{m_\chi^2}{m_A^2} \left( \frac{v_2^2}{v_1^2+v_2^2}\right)^2 \,,
\ee 
where $\sigma_{AN}$ is given by Eq.~(\ref{eq:VSI}).
Unlike in the case $v_1 \gg v_2$ (i.e. $m_{A^6} \gg m_A$), this cross section is significant. 
Note that $\sigma_{\chi N}$ is suppressed by the factor $m_\chi^2/m_A^2$ with respect to $\sigma_{A N}$.

In the presence of non-negligible scattering cross-sections for both DM components, the analysis of the Direct Detection limits is not straightforward. Such limits are normally 
given in terms of the DM--nucleon scattering cross-section as a function of the DM mass.
In our case, one should compare directly the experimental outcome, i.e.~the distribution of events with respect to the recoil energy, with the theoretical prediction
\be
\label{eq:energy_dist}
\frac{dN}{dE_R}=\sum_{i=\chi,A} f_i {\left(\frac{dN}{dE_R}\right)}_i \,.
\ee
Here $ f_i =\Omega_i/\Omega_{\rm tot}$ and
\bal
 {\left(\frac{dN}{dE_R}\right)}_i = \frac{\sigma_{i N}\rho_0}{2m_R^2m_i}F_i^2(E_R)\int_{v_\mathrm{min}(E_R)}^{\infty}\frac{f(v_i)}{v_i}dv_{i} \,,
\eal
with $\rho_0$ being the experimental value of the local DM density; $m_R$ is
the reduced mass of the DM--nucleus system, $F_i(E_R)^2$ is the form-factor due to the finite size of the nucleus (normalized to $F_i(0)^2=1 $), and $f(v_i)$ is the DM velocity
distribution in the detector frame. A detailed discussion of the Direct Detection limit
interpretation for multicomponent DM is given in~\cite{Profumo:2009tb,Dienes:2012cf}. 
Here we have adopted a simple approximate procedure. 
We have computed the total number of recoil events, obtained by integrating the distribution of Eq.~(\ref{eq:energy_dist})\footnote{The correct number of recoil events is actually given by the convolution of Eq.~(\ref{eq:energy_dist}) with a function accounting for the detector efficiency and finite energy resolution~\cite{Akrami:2010dn,Savage:2015xta}. Neglecting this function implies an overestimate of the number of recoil events for a given scattering cross-section. We have suitably chosen the limit number of events to partially compensate this effect.} over a suitable range of recoil energies and multiplied the result 
with the number of nuclei and the exposure time in a given experiment.
Given the design similarity between the LUX and XENON1T experiments, we have assumed 
the upper limit of 3 events
 for both (with two years of exposure time)
 \cite{Savage:2015xta}.
 This number takes into account the detector efficiency which is set to 1 in Micromegas.

It is seen from Fig.~\ref{fig:dm2} that the contribution from the pseudoscalar component tightens the limits from DD. Yet, the thermal DM relic density band
is still out of reach of LUX. The relevant Direct Detection suppression factors include a low value of $\sin\theta$, $m_\chi^2/m_A^2$ for 
a light $\chi$ component as well as a relatively small $\tilde g$ in
the domain of the broad resonance $m_\chi \sim {\cal O}(m_{h_2}/2)$.
We find that these factors are efficient enough to make the detection of a light
$\chi$ beyond the reach of XENON1T, while some regions with a heavier 
$\chi$ can be probed. This differs from the pure vector DM case considered in \cite{Gross:2015cwa}.

%=========================================================================
\subsection{Complementarity of direct and indirect detection}
%=========================================================================

In this subsection, we briefly explore 
 the possibility of observing one DM component in Indirect Detection (ID) experiments and the other one through Direct Detection. 

As seen in Eqs.~(\ref{eq:scalarxsection},\ref{eq:vectorxsection}), the pair annihilation cross-sections 
 are s-wave dominated and suffer no velocity suppression. Therefore,
both  the pseudoscalar and the vector DM components can potentially generate  an ID (photon)  signal from the $\bar b b, \bar t t, W^+ W^-, ZZ, hh$ final states. 
 However, as explained in the previous subsections, the vector component often annihilates into $\chi\chi$ most efficiently.
As a result, the ID signal would be suppressed and thus only the pseudoscalar component could potentially be detected.
This situation reverses in Direct Detection since the $\sigma_{\chi N}$
cross section is too small.

\begin{figure}
\begin{center}
\includegraphics[width=7.5 cm]{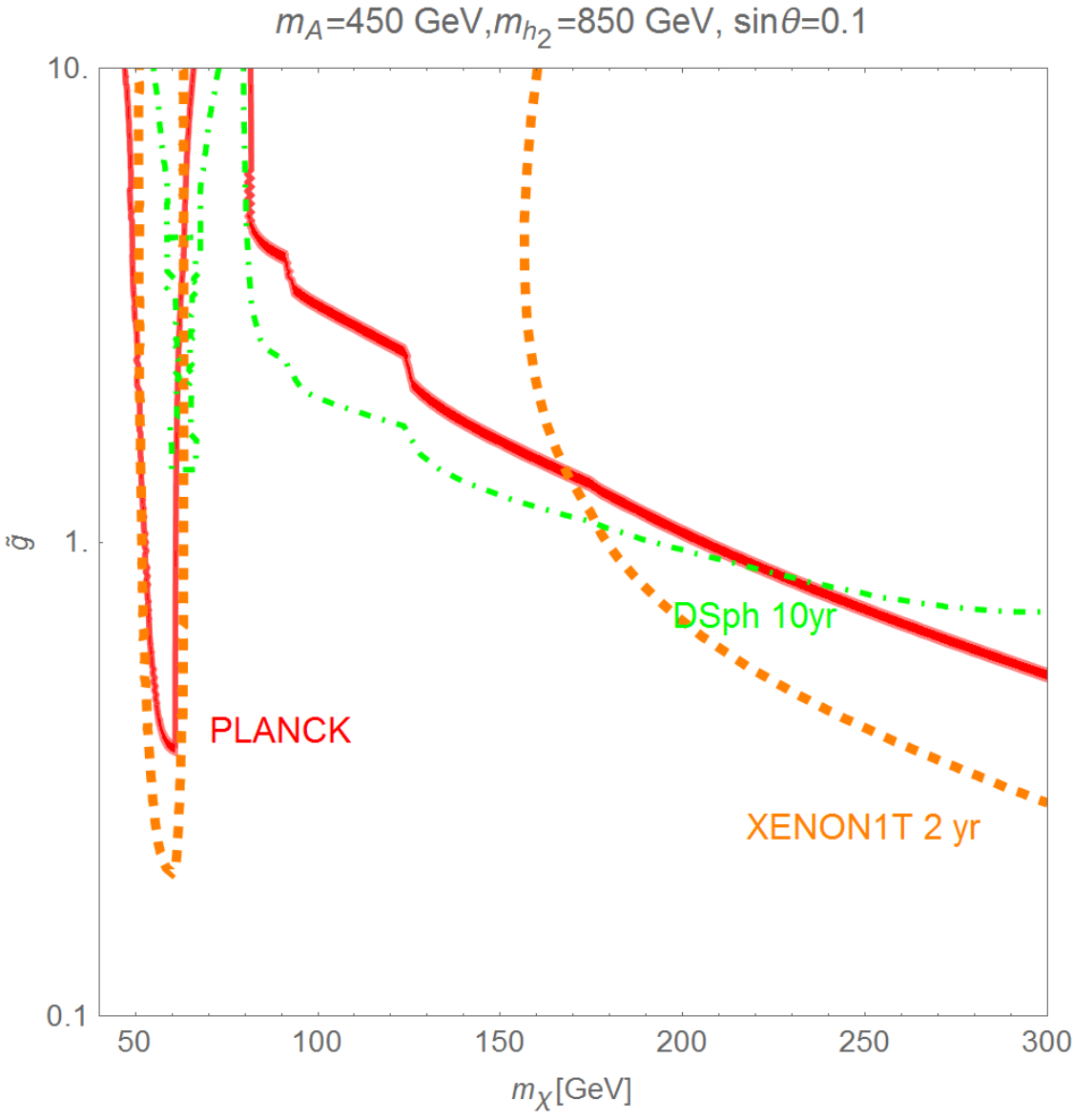}
\end{center}
\caption{Prospects of  detecting directly the vector DM component and indirectly  the pseudoscalar DM component.
The red band corresponds to the correct total DM relic density, the orange dashed line represents  projected DD limits from XENON1T with 
a 2 year exposure time 
and the dashed green line shows projected ID limits from FERMI~\cite{Ackermann:2013yva} with 
10 years of data taking~\cite{DphS10years}.}
\label{fig:pIDchi}
\end{figure}
While a detailed analysis is beyond the scope of this work, we illustrate this point with the following example (Fig.~\ref{fig:pIDchi}). 
We take $m_{h_2}=850~\GeV$, $m_A=450~\GeV$, $s_\theta=0.1$ and focus on the range $50-300~\GeV$ for the $\chi$ mass. 
These parameters are chosen in order to have a large pseudoscalar component since the ID rate scales with the square of the DM density.

We see that there are two small regions in Fig.~\ref{fig:pIDchi} where both ID and DD signals could  be detected. 
The first one is close to the $h_1$ resonance, i.e.~for $m_\chi \simeq 60~\GeV$.
This region may in fact  be compatible with the Galactic Center gamma-ray excess~\cite{GCexcess} although reproducing it in vicinity of the s-channel resonances is in general rather contrived~\cite{NMSSMGC}.
The second region corresponds to masses $m_\chi \sim 170-230\,\mbox{GeV}$. In this region, the density fraction of the vectorial DM component is very low, for example, $f_A \approx 0.02$ at $m_\chi=170\,\mbox{GeV}$. This is compensated by the high DD cross section since
the correct relic density requires  $\tilde{g} \simeq 1$.
Specifically, 
for $m_A=450\,\mbox{GeV}$ and $m_\chi=170\,\mbox{GeV}$, 
one has $f_A \sigma_{NA} \approx 7.5 \times 10^{-47}\,{\mbox{cm}}^2$. 

A complication here is that it is very difficult to prove that 
DD and ID signals come from particles with different masses.
One obstacle is the large uncertainty (hundreds of GeV) in the DM
mass determination through  Direct Detection  (cf. eg.~\cite{Strege:2012kv}). This stems from the very weak dependence   of the spectrum of recoil events on the DM mass (for heavy DM). 
Thus, in practice it would be challenging to prove that Dark Matter is indeed multicomponent. 

Further information which can help deciphering 
 the DM composition would be provided by collider experiments.
 In particular, in certain kinematic regimes, e.g. $m_{h_2} > 2 m_{A}$
 and $m_{h_2}< 300$ GeV, 
the LHC monojet events with missing energy will be able to probe the hidden sector gauge coupling in the range ${\cal O}(10^{-1})- {\cal O}(1)$ \cite{Kim:2015hda}. 
Similar constraints are obtained in Vector Boson Fusion  \cite{Chen:2015dea} (see also \cite{DiFranzo:2015nli}).
Other channels can provide further probes, which will be studied elsewhere.   

%=========================================================================
%=========================================================================
\section{Conclusions}
%=========================================================================
%=========================================================================

We have studied a simple UV complete set--up which entails naturally multicomponent Dark Matter with spin-1 and spin-0 constituents. 
The symmetry that stabilizes DM is not put in by hand, but is instead 
inherent in the Yang--Mills system. 
The model belongs to the Higgs portal category with the hidden sector consisting
of SU(3) Yang--Mills fields as well as the minimal Higgs content to break this symmetry completely. 
Upon spontaneous gauge symmetry breaking, the system retains a global
$U(1) \times Z_2^\prime$ symmetry (assuming unbroken {\it CP} in the hidden sector).  We focus on its discrete subgroup 
 $Z_2 \times Z_2^\prime$ which can be regarded as a DM stabilizer   making the lightest vector fields and a pseudoscalar stable. These play the role of multicomponent Dark Matter.

Even though the theory is rather simple in the UV, the DM phenomenology is very rich offering a number of qualitatively different parametric regimes. For instance, the ``dark annihilation'' channel, where the heavier DM component pair--annihilates into the lighter component, can play an important role. Dark Matter can be mostly spin-1, mostly spin-0 or mixed.
 An attractive feature of 
the model is that the Direct DM Detection rate is suppressed as long as
the SM Higgs mixes predominantly with a single scalar of the hidden sector. This phenomenon is qualitatively different from the known
DD suppression mechanisms. We find that in many regions of parameter space, the Direct Detection rate is well below the LUX2016 
(and sometimes XENON1T) constraint
while still consistent with the thermal WIMP paradigm.

This shows, in particular, that the Higgs portal Dark Matter framework
offers a number of viable options and the WIMP paradigm is not necessarily in crisis.

\vspace{10pt}
{\bf Acknowledgements} 

The authors thank A.~Pukhov for his help with the package Micromegas.
G.A. acknowledges support from the ERC advanced grant {\it Higgs@LHC}.
C.G. is grateful to the {\it Mainz Institute for Theoretical Physics (MITP)}, where a part of this work was done, for its hospitality and partial financial support.
C.G. and O.L. acknowledge support from the Academy of Finland, project {\it The Higgs Boson and the Cosmos}. 
This project has received funding from the European Union's Horizon 2020 research and innovation programme under the Marie Sklodowska-Curie grant agreement No 674896.
S.P. is supported by the National Science Centre, Poland, under research grants
DEC-2014/15/B/ST2/02157, DEC-2015/19/B/ST2/02848
and DEC-2015/18/M/ST2/00054.
T.T. acknowledges support from P2IO Excellence Laboratory (LABEX).

{}


\begin{thebibliography}{}

\bibitem{PLANCK} P.~A.~R.~Ade {\it et al.} [Planck Collaboration], arXiv:1303.5076 [astro-ph.CO].


%\cite{Silveira:1985rk}
\bibitem{Silveira:1985rk} 
  V.~Silveira and A.~Zee,
  %``Scalar Phantoms,''
  Phys.\ Lett.\  {\bf 161B}, 136 (1985).
  %doi:10.1016/0370-2693(85)90624-0
  %%CITATION = doi:10.1016/0370-2693(85)90624-0;%%


 %\cite{Patt:2006fw}
\bibitem{Patt:2006fw}
 B.~Patt and F.~Wilczek,
 %``Higgs-field portal into hidden sectors,''
 hep-ph/0605188.
 %%CITATION = HEP-PH/0605188;%%
 
 %\cite{MarchRussell:2008yu}
\bibitem{MarchRussell:2008yu}
 J.~March-Russell, S.~M.~West, D.~Cumberbatch and D.~Hooper,
 %``Heavy Dark Matter Through the Higgs Portal,''
 JHEP {\bf 0807} (2008) 058
% doi:10.1088/1126-6708/2008/07/058
 [arXiv:0801.3440 [hep-ph]].
 
 %\cite{Andreas:2008xy}
\bibitem{Andreas:2008xy}
 S.~Andreas, T.~Hambye and M.~H.~G.~Tytgat,
 %``WIMP dark matter, Higgs exchange and DAMA,''
 JCAP {\bf 0810} (2008) 034
 %doi:10.1088/1475-7516/2008/10/034
 [arXiv:0808.0255 [hep-ph]].
 
%\cite{Hambye:2008bq}
\bibitem{Hambye:2008bq} 
  T.~Hambye,
  %``Hidden vector dark matter,''
  JHEP {\bf 0901}, 028 (2009)
 % doi:10.1088/1126-6708/2009/01/028
  [arXiv:0811.0172 [hep-ph]].
  %%CITATION = doi:10.1088/1126-6708/2009/01/028;%%
  %122 citations counted in INSPIRE as of 04 Nov 2016
 
 
 
 %\cite{Kanemura:2010sh}
\bibitem{Kanemura:2010sh}
 S.~Kanemura, S.~Matsumoto, T.~Nabeshima and N.~Okada,
 %``Can WIMP Dark Matter overcome the Nightmare Scenario?,''
 Phys.\ Rev.\ D {\bf 82} (2010) 055026
 % doi:10.1103/PhysRevD.82.055026
 [arXiv:1005.5651 [hep-ph]].
 %%CITATION = doi:10.1103/PhysRevD.82.055026;%%
 %129 citations counted in INSPIRE as of 09 Sep 2016
 
 %\cite{Lebedev:2011iq}
\bibitem{Lebedev:2011iq}
 O.~Lebedev, H.~M.~Lee and Y.~Mambrini,
 %``Vector Higgs-portal dark matter and the invisible Higgs,''
 Phys.\ Lett.\ B {\bf 707} (2012) 570
% doi:10.1016/j.physletb.2012.01.029
 [arXiv:1111.4482 [hep-ph]].
 
 
  %\cite{Djouadi:2011aa}
\bibitem{Djouadi:2011aa} 
  A.~Djouadi, O.~Lebedev, Y.~Mambrini and J.~Quevillon,
  %``Implications of LHC searches for Higgs--portal dark matter,''
  Phys.\ Lett.\ B {\bf 709}, 65 (2012)
%  doi:10.1016/j.physletb.2012.01.062
  [arXiv:1112.3299 [hep-ph]].
  %%CITATION = doi:10.1016/j.physletb.2012.01.062;%%
  %259 citations counted in INSPIRE as of 31 Oct 2016
 

 %\cite{Kim:2006af}
\bibitem{Kim:2006af}
 Y.~G.~Kim and K.~Y.~Lee,
 %``The Minimal model of fermionic dark matter,''
 Phys.\ Rev.\ D {\bf 75} (2007) 115012
% doi:10.1103/PhysRevD.75.115012
 [hep-ph/0611069].
 
 
 
 %\cite{LopezHonorez:2012kv}
\bibitem{LopezHonorez:2012kv}
 L.~Lopez-Honorez, T.~Schwetz and J.~Zupan,
 %``Higgs portal, fermionic dark matter, and a Standard Model like Higgs at 125 GeV,''
 Phys.\ Lett.\ B {\bf 716} (2012) 179
 %doi:10.1016/j.physletb.2012.07.017
 [arXiv:1203.2064 [hep-ph]].
 
 
%\cite{Farzan:2012hh}
\bibitem{Farzan:2012hh}
 Y.~Farzan and A.~R.~Akbarieh,
 %``VDM: A model for Vector Dark Matter,''
 JCAP {\bf 1210} (2012) 026
 [arXiv:1207.4272 [hep-ph]].
 %%CITATION = ARXIV:1207.4272;%%
 
%\cite{Baek:2012se}
\bibitem{Baek:2012se} 
  S.~Baek, P.~Ko, W.~I.~Park and E.~Senaha,
  %``Higgs Portal Vector Dark Matter : Revisited,''
  JHEP {\bf 1305}, 036 (2013)
 % doi:10.1007/JHEP05(2013)036
  [arXiv:1212.2131 [hep-ph]].
  %%CITATION = doi:10.1007/JHEP05(2013)036;%%
  %60 citations counted in INSPIRE as of 31 Oct 2016
 
%\cite{Duch:2015jta}
\bibitem{Duch:2015jta} 
  M.~Duch, B.~Grzadkowski and M.~McGarrie,
  %``A stable Higgs portal with vector dark matter,''
  JHEP {\bf 1509}, 162 (2015)
  %doi:10.1007/JHEP09(2015)162
  [arXiv:1506.08805 [hep-ph]].
  %%CITATION = doi:10.1007/JHEP09(2015)162;%%
  %10 citations counted in INSPIRE as of 31 Oct 2016
  
  
  
  %\bibitem{Akerib:2013tjd} 
\bibitem{LUX} D.~S.~Akerib {\it et al.} [LUX Collaboration], arXiv:1310.8214 [astro-ph.CO]; D.~S.~Akerib {\it et al.} [LUX Collaboration], arXiv:1512.03506 [astro-ph.CO].


%\cite{Tan:2016zwf}
\bibitem{Tan:2016zwf} 
  A.~Tan {\it et al.} [PandaX-II Collaboration],
  %``Dark Matter Results from First 98.7 Days of Data from the PandaX-II Experiment,''
  Phys.\ Rev.\ Lett.\  {\bf 117}, no. 12, 121303 (2016)
 % doi:10.1103/PhysRevLett.117.121303
  [arXiv:1607.07400 [hep-ex]].
  %%CITATION = doi:10.1103/PhysRevLett.117.121303;%%
  %49 citations counted in INSPIRE as of 04 Nov 2016



%\cite{Belanger:2011ww}
\bibitem{Belanger:2011ww}
 G.~Belanger and J.~C.~Park,
 %``Assisted freeze-out,''
 JCAP {\bf 1203} (2012) 038
 doi:10.1088/1475-7516/2012/03/038
 [arXiv:1112.4491 [hep-ph]].
 %%CITATION = doi:10.1088/1475-7516/2012/03/038;%%
 %26 citations counted in INSPIRE as of 05 Jul 2016
 
 
 
%\cite{Dudas:2014ixa}
\bibitem{Dudas:2014ixa}
 E.~Dudas, L.~Heurtier and Y.~Mambrini,
 %``Generating X-ray lines from annihilating dark matter,''
 Phys.\ Rev.\ D {\bf 90} (2014) 035002
 %doi:10.1103/PhysRevD.90.035002
 [arXiv:1404.1927 [hep-ph]].
 %%CITATION = doi:10.1103/PhysRevD.90.035002;%%
 %48 citations counted in INSPIRE as of 13 Jul 2016
 


%\cite{Dienes:2012cf}
\bibitem{Dienes:2012cf}
 K.~R.~Dienes, J.~Kumar and B.~Thomas,
 %``Direct Detection of Dynamical Dark Matter,''
 Phys.\ Rev.\ D {\bf 86} (2012) 055016
 %doi:10.1103/PhysRevD.86.055016
 [arXiv:1208.0336 [hep-ph]].
 %%CITATION = doi:10.1103/PhysRevD.86.055016;%%
 %17 citations counted in INSPIRE as of 05 Jul 2016

%\cite{Profumo:2009tb}
\bibitem{Profumo:2009tb}
 S.~Profumo, K.~Sigurdson and L.~Ubaldi,
 %``Can we discover multi-component WIMP dark matter?,''
 JCAP {\bf 0912} (2009) 016
 %doi:10.1088/1475-7516/2009/12/016
 [arXiv:0907.4374 [hep-ph]].
 %%CITATION = doi:10.1088/1475-7516/2009/12/016;%%
 %28 citations counted in INSPIRE as of 05 Jul 2016
	

 
 %\cite{Esch:2014jpa}
\bibitem{Esch:2014jpa}
 S.~Esch, M.~Klasen and C.~E.~Yaguna,
 %``A minimal model for two-component dark matter,''
 JHEP {\bf 1409} (2014) 108
 %doi:10.1007/JHEP09(2014)108
 [arXiv:1406.0617 [hep-ph]].
 %%CITATION = doi:10.1007/JHEP09(2014)108;%%
 %8 citations counted in INSPIRE as of 05 Jul 2016
 
 %\cite{Boddy:2014yra}
\bibitem{Boddy:2014yra} 
  K.~K.~Boddy, J.~L.~Feng, M.~Kaplinghat and T.~M.~P.~Tait,
  %``Self-Interacting Dark Matter from a Non-Abelian Hidden Sector,''
  Phys.\ Rev.\ D {\bf 89}, no. 11, 115017 (2014)
  %doi:10.1103/PhysRevD.89.115017
  [arXiv:1402.3629 [hep-ph]].
  %%CITATION = doi:10.1103/PhysRevD.89.115017;%%
  %58 citations counted in INSPIRE as of 31 Oct 2016

 
%\cite{Klasen:2016qux}
\bibitem{Klasen:2016qux}
  M.~Klasen, F.~Lyonnet and F.~S.~Queiroz,
  %``NLO+NLL Collider Bounds, Dirac Fermion and Scalar Dark Matter in the B-L Model,''
  arXiv:1607.06468 [hep-ph].
  %%CITATION = ARXIV:1607.06468;%%
  %4 citations counted in INSPIRE as of 28 Oct 20 
 
%\cite{DiFranzo:2016uzc}
\bibitem{DiFranzo:2016uzc} 
  A.~DiFranzo and G.~Mohlabeng,
  %``Multi-component Dark Matter through a Radiative Higgs Portal,''
  arXiv:1610.07606 [hep-ph].
  %%CITATION = ARXIV:1610.07606;%% 
 
 
 %\cite{Gross:2015cwa}
\bibitem{Gross:2015cwa}
 C.~Gross, O.~Lebedev and Y.~Mambrini,
 %``Non-Abelian gauge fields as dark matter,''
 JHEP {\bf 1508} (2015) 158
 [arXiv:1505.07480 [hep-ph]].
 %%CITATION = ARXIV:1505.07480;%%

 
%\cite{Karam:2015jta}
\bibitem{Karam:2015jta} 
  A.~Karam and K.~Tamvakis,
  %``Dark matter and neutrino masses from a scale-invariant multi-Higgs portal,''
  Phys.\ Rev.\ D {\bf 92}, no. 7, 075010 (2015)
  %doi:10.1103/PhysRevD.92.075010
  [arXiv:1508.03031 [hep-ph]].
  %%CITATION = doi:10.1103/PhysRevD.92.075010;%%
  %21 citations counted in INSPIRE as of 04 Nov 2016 
 

 %\cite{Khoze:2016zfi}
\bibitem{Khoze:2016zfi} 
 V.~V.~Khoze and A.~D.~Plascencia,
 %``Dark Matter and Leptogenesis Linked by Classical Scale Invariance,''
 arXiv:1605.06834 [hep-ph].
 %%CITATION = ARXIV:1605.06834;%%
 
 

%\cite{Karam:2016rsz}
\bibitem{Karam:2016rsz}
 A.~Karam and K.~Tamvakis,
 %``Dark Matter from a Classically Scale-Invariant $SU(3)_X$,''
 arXiv:1607.01001 [hep-ph].
 %%CITATION = ARXIV:1607.01001;%%
 
 	
%\cite{Bae:2015rra}
\bibitem{Bae:2015rra}
 K.~J.~Bae, H.~Baer, A.~Lessa and H.~Serce,
 %``Mixed axion-wino dark matter,''
 Front.\ in Phys.\ {\bf 3} (2015) 49
 %doi:10.3389/fphy.2015.00049
 [arXiv:1502.07198 [hep-ph]].
 %%CITATION = doi:10.3389/fphy.2015.00049;%%
 %4 citations counted in INSPIRE as of 23 Jul 2016



%\cite{Badziak:2015qca}
\bibitem{Badziak:2015qca}
 M.~Badziak, A.~Delgado, M.~Olechowski, S.~Pokorski and K.~Sakurai,
 %``Detecting underabundant neutralinos,''
 JHEP {\bf 1511} (2015) 053
 %doi:10.1007/JHEP11(2015)053
 [arXiv:1506.07177 [hep-ph]].
 %%CITATION = doi:10.1007/JHEP11(2015)053;%%
 %12 citations counted in INSPIRE as of 23 Jul 2016
 
 
%\cite{Aprile:2015uzo}
\bibitem{Aprile:2015uzo}
 E.~Aprile {\it et al.} [XENON Collaboration],
 %``Physics reach of the XENON1T dark matter experiment,''
 JCAP {\bf 1604} (2016) no.04, 027
 %doi:10.1088/1475-7516/2016/04/027
 [arXiv:1512.07501 [physics.ins-det]].
 %%CITATION = doi:10.1088/1475-7516/2016/04/027;%%
 %48 citations counted in INSPIRE as of 23 Jul 2016
 
 
 %\cite{Falkowski:2015iwa}
\bibitem{Falkowski:2015iwa}
 A.~Falkowski, C.~Gross and O.~Lebedev,
 %``A second Higgs from the Higgs portal,''
 JHEP {\bf 1505} (2015) 057
 %doi:10.1007/JHEP05(2015)057
 [arXiv:1502.01361 [hep-ph]].
 %%CITATION = doi:10.1007/JHEP05(2015)057;%%
 %51 citations counted in INSPIRE as of 30 Jul 2016

	
	

%\cite{Aoki:2012ub}
\bibitem{Aoki:2012ub}
 M.~Aoki, M.~Duerr, J.~Kubo and H.~Takano,
 %``Multi-Component Dark Matter Systems and Their Observation Prospects,''
 Phys.\ Rev.\ D {\bf 86} (2012) 076015
 %doi:10.1103/PhysRevD.86.076015
 [arXiv:1207.3318 [hep-ph]].
 %%CITATION = doi:10.1103/PhysRevD.86.076015;%%
 %36 citations counted in INSPIRE as of 05 Jul 2016
	
	
%\cite{D'Eramo:2010ep}
\bibitem{D'Eramo:2010ep}
 F.~D'Eramo and J.~Thaler,
 %``Semi-annihilation of Dark Matter,''
 JHEP {\bf 1006} (2010) 109
 %doi:10.1007/JHEP06(2010)109
 [arXiv:1003.5912 [hep-ph]].
 %%CITATION = doi:10.1007/JHEP06(2010)109;%%
 %50 citations counted in INSPIRE as of 21 Jul 2016
	
%\cite{Belanger:2012vp}
\bibitem{Belanger:2012vp}
 G.~Belanger, K.~Kannike, A.~Pukhov and M.~Raidal,
 %``Impact of semi-annihilations on dark matter phenomenology - an example of Z_N symmetric scalar dark matter,''
 JCAP {\bf 1204} (2012) 010
 %doi:10.1088/1475-7516/2012/04/010
 [arXiv:1202.2962 [hep-ph]].
 %%CITATION = doi:10.1088/1475-7516/2012/04/010;%%
 %42 citations counted in INSPIRE as of 21 Jul 2016
 
 
 %\cite{Edsjo:1997bg}
\bibitem{Edsjo:1997bg}
 J.~Edsjo and P.~Gondolo,
 %``Neutralino relic density including coannihilations,''
 Phys.\ Rev.\ D {\bf 56} (1997) 1879
 %doi:10.1103/PhysRevD.56.1879
 [hep-ph/9704361].
 %%CITATION = doi:10.1103/PhysRevD.56.1879;%%
 %444 citations counted in INSPIRE as of 23 Jul 2016

%\cite{Gondolo:1990dk}
\bibitem{Gondolo:1990dk}
 P.~Gondolo and G.~Gelmini,
 %``Cosmic abundances of stable particles: Improved analysis,''
 Nucl.\ Phys.\ B {\bf 360} (1991) 145.
 %doi:10.1016/0550-3213(91)90438-4
 %%CITATION = doi:10.1016/0550-3213(91)90438-4;%%
 %701 citations counted in INSPIRE as of 24 Jul 2016
 
 %\cite{Jungman:1995df}
\bibitem{Jungman:1995df}
  G.~Jungman, M.~Kamionkowski and K.~Griest,
  %``Supersymmetric dark matter,''
  Phys.\ Rept.\  {\bf 267} (1996) 195
%  doi:10.1016/0370-1573(95)00058-5
  [hep-ph/9506380].
 
 
 %\cite{Belanger:2014vza}
\bibitem{Belanger:2014vza} 
 G.~Belanger, F.~Boudjema, A.~Pukhov and A.~Semenov,
 %``micrOMEGAs4.1: two dark matter candidates,''
 Comput.\ Phys.\ Commun.\ {\bf 192}, 322 (2015)
 %doi:10.1016/j.cpc.2015.03.003
 [arXiv:1407.6129 [hep-ph]].
 %%CITATION = doi:10.1016/j.cpc.2015.03.003;%%
 %49 citations counted in INSPIRE as of 08 Jan 2016
 
 %\cite{Akerib:2015rjg}
\bibitem{Akerib:2015rjg}
 D.~S.~Akerib {\it et al.} [LUX Collaboration],
 %``Improved Limits on Scattering of Weakly Interacting Massive Particles from Reanalysis of 2013 LUX Data,''
 Phys.\ Rev.\ Lett.\ {\bf 116} (2016) no.16, 161301
 %doi:10.1103/PhysRevLett.116.161301
 [arXiv:1512.03506 [astro-ph.CO]].
 %%CITATION = doi:10.1103/PhysRevLett.116.161301;%%
 %124 citations counted in INSPIRE as of 24 Jul 2016



%\cite{Boehm:2014hva}
\bibitem{Boehm:2014hva}
 C.~Boehm, M.~J.~Dolan, C.~McCabe, M.~Spannowsky and C.~J.~Wallace,
 %``Extended gamma-ray emission from Coy Dark Matter,''
 JCAP {\bf 1405} (2014) 009
 %doi:10.1088/1475-7516/2014/05/009
 [arXiv:1401.6458 [hep-ph]].
 %%CITATION = doi:10.1088/1475-7516/2014/05/009;%%
 %95 citations counted in INSPIRE as of 24 Jul 2016

%\cite{Lebedev:2014bba}
\bibitem{Lebedev:2014bba} 
 O.~Lebedev and Y.~Mambrini,
 %``Axial dark matter: The case for an invisible $Z′$,''
 Phys.\ Lett.\ B {\bf 734}, 350 (2014)
 %doi:10.1016/j.physletb.2014.05.025
 [arXiv:1403.4837 [hep-ph]].
 %%CITATION = doi:10.1016/j.physletb.2014.05.025;%%
 %37 citations counted in INSPIRE as of 24 Oct 2016
 
 
 %\cite{Crivellin:2013ipa}
\bibitem{Crivellin:2013ipa}
  A.~Crivellin, M.~Hoferichter and M.~Procura,
  %``Accurate evaluation of hadronic uncertainties in spin-independent WIMP-nucleon scattering: Disentangling two- and three-flavor effects,''
  Phys.\ Rev.\ D {\bf 89} (2014) 054021
%  doi:10.1103/PhysRevD.89.054021
  [arXiv:1312.4951 [hep-ph]].
 
 
 
 %\cite{Savage:2015xta}
\bibitem{Savage:2015xta}
 C.~Savage, A.~Scaffidi, M.~White and A.~G.~Williams,
 %``LUX likelihood and limits on spin-independent and spin-dependent WIMP couplings with LUXCalc,''
 Phys.\ Rev.\ D {\bf 92} (2015) no.10, 103519
 %doi:10.1103/PhysRevD.92.103519
 [arXiv:1502.02667 [hep-ph]].
 %%CITATION = doi:10.1103/PhysRevD.92.103519;%%
 %20 citations counted in INSPIRE as of 13 Jul 2016
 
 
  
 %\cite{Akrami:2010dn}
\bibitem{Akrami:2010dn}
 Y.~Akrami, C.~Savage, P.~Scott, J.~Conrad and J.~Edsjo,
 %``How well will ton-scale dark matter direct detection experiments constrain minimal supersymmetry?,''
 JCAP {\bf 1104} (2011) 012
 %doi:10.1088/1475-7516/2011/04/012
 [arXiv:1011.4318 [astro-ph.CO]].
 %%CITATION = doi:10.1088/1475-7516/2011/04/012;%%
 %33 citations counted in INSPIRE as of 13 Jul 2016
 
 
 
%
\bibitem{Ackermann:2013yva}
 M.~Ackermann {\it et al.} [Fermi-LAT Collaboration],
 %``Dark matter constraints from observations of 25 Milky Way satellite galaxies with the Fermi Large Area Telescope,''
 Phys.\ Rev.\ D {\bf 89} (2014) 4, 042001
 [arXiv:1310.0828 [astro-ph.HE]].
 %%CITATION = ARXIV:1310.0828;%%
 %92 citations counted in INSPIRE as of 16 Oct 2014
 
 
 \bibitem{DphS10years}
Private communication from German Gomez.


%\cite{Hooper:2010mq}
\bibitem{GCexcess}
 D.~Hooper and L.~Goodenough,
 %``Dark Matter Annihilation in The Galactic Center As Seen by the Fermi Gamma Ray Space Telescope,''
 Phys.\ Lett.\ B {\bf 697} (2011) 412
 [arXiv:1010.2752 [hep-ph]];
 %%CITATION = ARXIV:1010.2752;%%
 %198 citations counted in INSPIRE as of 16 Oct 2014	
	%\cite{Hooper:2011ti}
%\bibitem{Hooper:2011ti}
 D.~Hooper and T.~Linden,
 %``On The Origin Of The Gamma Rays From The Galactic Center,''
 Phys.\ Rev.\ D {\bf 84} (2011) 123005
 [arXiv:1110.0006 [astro-ph.HE]];
 %%CITATION = ARXIV:1110.0006;%%
 %163 citations counted in INSPIRE as of 16 Oct 2014
%\cite{Abazajian:2012pn}
%\bibitem{Abazajian:2012pn}
 K.~N.~Abazajian and M.~Kaplinghat,
 %``Detection of a Gamma-Ray Source in the Galactic Center Consistent with Extended Emission from Dark Matter Annihilation and Concentrated Astrophysical Emission,''
 Phys.\ Rev.\ D {\bf 86} (2012) 083511
 [arXiv:1207.6047 [astro-ph.HE]];
 %%CITATION = ARXIV:1207.6047;%%
 %109 citations counted in INSPIRE as of 16 Oct 2014
	%\cite{Gordon:2013vta}
%\bibitem{Gordon:2013vta}
 C.~Gordon and O.~Macias,
 %``Dark Matter and Pulsar Model Constraints from Galactic Center Fermi-LAT Gamma Ray Observations,''
 Phys.\ Rev.\ D {\bf 88} (2013) 083521
 [arXiv:1306.5725 [astro-ph.HE]].
 %%CITATION = ARXIV:1306.5725;%%
 %68 citations counted in INSPIRE as of 16 Oct 2014
%\cite{Abazajian:2014fta}
%\bibitem{Abazajian:2014fta}
 K.~N.~Abazajian, N.~Canac, S.~Horiuchi and M.~Kaplinghat,
 %``Astrophysical and Dark Matter Interpretations of Extended Gamma-Ray Emission from the Galactic Center,''
 Phys.\ Rev.\ D {\bf 90} (2014) 023526
 [arXiv:1402.4090 [astro-ph.HE]];
 %%CITATION = ARXIV:1402.4090;%%
 %55 citations counted in INSPIRE as of 16 Oct 2014	
 F.~Calore, I.~Cholis and C.~Weniger,
 %``Background model systematics for the Fermi GeV excess,''
 arXiv:1409.0042 [astro-ph.CO].
 %%CITATION = ARXIV:1409.0042;%%
 %6 citations counted in INSPIRE as of 14 Oct 2014
 
 
 %\cite{Cheung:2014lqa}
\bibitem{NMSSMGC}
 C.~Cheung, M.~Papucci, D.~Sanford, N.~R.~Shah and K.~M.~Zurek,
 %``NMSSM Interpretation of the Galactic Center Excess,''
 Phys.\ Rev.\ D {\bf 90} (2014) 075011
 [arXiv:1406.6372 [hep-ph]].
 %%CITATION = ARXIV:1406.6372;%%
 %13 citations counted in INSPIRE as of 29 Oct 2014
 


%\cite{Strege:2012kv}
\bibitem{Strege:2012kv}
 C.~Strege, R.~Trotta, G.~Bertone, A.~H.~G.~Peter and P.~Scott,
 %``Fundamental statistical limitations of future dark matter direct detection experiments,''
 Phys.\ Rev.\ D {\bf 86} (2012) 023507
 %doi:10.1103/PhysRevD.86.023507
 [arXiv:1201.3631 [hep-ph]].
 %%CITATION = doi:10.1103/PhysRevD.86.023507;%%
 %32 citations counted in INSPIRE as of 24 Jul 2016
	

%\cite{Kim:2015hda}
\bibitem{Kim:2015hda} 
  J.~S.~Kim, O.~Lebedev and D.~Schmeier,
  %``Higgsophilic gauge bosons and monojets at the LHC,''
  JHEP {\bf 1511}, 128 (2015)
  %doi:10.1007/JHEP11(2015)128
  [arXiv:1507.08673 [hep-ph]].
  %%CITATION = doi:10.1007/JHEP11(2015)128;%%
  %4 citations counted in INSPIRE as of 31 Oct 2016

%\cite{Chen:2015dea}
\bibitem{Chen:2015dea} 
  C.~H.~Chen and T.~Nomura,
  %``Searching for vector dark matter via Higgs portal at the LHC,''
  Phys.\ Rev.\ D {\bf 93}, no. 7, 074019 (2016)
  %doi:10.1103/PhysRevD.93.074019
  [arXiv:1507.00886 [hep-ph]].
  %%CITATION = doi:10.1103/PhysRevD.93.074019;%%
  %6 citations counted in INSPIRE as of 31 Oct 2016

%\cite{DiFranzo:2015nli}
\bibitem{DiFranzo:2015nli} 
  A.~DiFranzo, P.~J.~Fox and T.~M.~P.~Tait,
  %``Vector Dark Matter through a Radiative Higgs Portal,''
  JHEP {\bf 1604}, 135 (2016)
 % doi:10.1007/JHEP04(2016)135
  [arXiv:1512.06853 [hep-ph]].
  %%CITATION = doi:10.1007/JHEP04(2016)135;%%
  %4 citations counted in INSPIRE as of 31 Oct 2016


\end{thebibliography}
\end{document}